\documentclass[12pt]{article}

\usepackage{xr}
\usepackage{float}
\usepackage{graphicx}
\usepackage{caption}
\usepackage{xcolor}
\usepackage{amsmath}
\usepackage{amssymb}
\usepackage{booktabs}
\usepackage{array}
\usepackage{multirow}
\usepackage{setspace}
\usepackage{enumitem}
\usepackage{siunitx}
\usepackage{authblk}
\usepackage{subcaption}
\usepackage[numbers,sort&compress]{natbib}
\usepackage[a4paper,margin=1in]{geometry}

\usepackage{xr-hyper}
\usepackage{hyperref}

\hypersetup{
    colorlinks=true,
    linkcolor=blue!70!black,
    citecolor=blue!70!black,
    urlcolor=blue!70!black
}

\begin{document}

\title{\textbf{Atomistic Indicators of the Ductile-to-Brittle Transition in Polycrystalline Tungsten: Temperature and Rhenium Effects on Crack-Tip Plasticity}}

\author{Divyesh A. Mistry}
\author{Avik Mahata\thanks{Corresponding author: \href{mailto:mahataa@merrimack.edu}{mahataa@merrimack.edu}}}

\affil{Department of Mechanical and Electrical Engineering, Merrimack College, North Andover, MA, USA}

\date{}

\maketitle
\begin{abstract}
The fracture response of body-centered-cubic tungsten is governed by the competition between crack-tip instability and dislocation-mediated plastic accommodation. Here, molecular dynamics (MD) simulations are used to examine edge-cracked polycrystalline W and W--Re under displacement-controlled Mode-I tension over 300--1800~K at a strain rate of \(5\times10^{8}~\mathrm{s}^{-1}\). Heating pure W reduces the instability stress, effective specimen stiffness, and volumetric work density accumulated prior to mechanical instability, whereas the corresponding instability strain remains comparatively scattered. An energy-based analysis of the pre-instability work defines a normalized work-loss metric, whose sigmoid fit yields a broad crossover marker, \(T_{\mathrm{MD}}^{*}\approx870~\mathrm{K}\), characteristic of the simulated geometry and high-rate loading conditions. Among the available W--Re compositions, W--10Re exhibits greater pre-instability deformation and work accumulation than pure W across most temperatures, while the maximum-stress response remains strongly dependent on temperature and composition. The accompanying changes in retained dislocation character and near-tip activity indicate that Re modifies the crack-tip plastic-accommodation pathway rather than acting through a simple strength increment. Analysis at common mechanical and temporal states further shows that Re alters the retained \(\frac{1}{2}\langle111\rangle\) line character and the temporal distribution of near-tip dislocation activity without uniformly increasing the local dislocation population. These coupled mechanical and defect-structure changes provide atomistic indicators of a high-rate ductile-to-brittle transition (DBT), but do not constitute a direct prediction of an experimental ductile-to-brittle transition temperature (DBTT). A separate OpenDiS/pydis calculation demonstrates source-like bow-out and heterogeneous dislocation-network development in crack-free polycrystalline W, providing representative mesoscale context without implying direct quantitative MD--DDD coupling.
\end{abstract}

\noindent\textbf{Keywords:} ductile-to-brittle transition; tungsten; tungsten--rhenium alloys; crack-tip plasticity; molecular dynamics; dislocation dynamics

\section{Introduction}
\label{sec:introduction}
Tungsten is a leading candidate plasma-facing and high-heat-flux material for fusion energy systems because of its exceptionally high melting point, high elastic modulus, high-temperature strength, thermal conductivity, low sputtering yield, and low hydrogen-isotope retention \cite{lassner1999,rieth2013,philipps2011,federici2003,wurster2013,ueda2014}. These properties make W attractive for divertor and first-wall environments, where materials must tolerate severe thermal loading, plasma exposure, and neutron irradiation. However, the practical use of tungsten is limited by intrinsic brittleness at low and intermediate temperatures, a high ductile-to-brittle transition temperature (DBTT), recrystallization-induced embrittlement, and irradiation-induced hardening \cite{wronski1965ductile,lassila1991ductile,zinkle2012,riethImpact2018,gumbsch2003,reiser2018mechanisms,reiser2018methods}. Improving the fracture resistance of W while retaining its high-temperature stability therefore remains a central challenge for fusion-relevant alloy design.

The ductile-to-brittle response of body-centered cubic (BCC) tungsten is controlled by the competition between cleavage crack advance and dislocation-mediated plastic relaxation near the crack tip. In BCC metals, plastic flow is strongly influenced by the mobility of $1/2\langle111\rangle$ screw dislocations, whose non-planar core
structure produces a high Peierls barrier and strong temperature dependence of glide \cite{vitek1968,cereceda2015linking,itakura2012,weinberger2013,srivastava2020}. At low temperature, limited dislocation mobility prevents sufficient crack-tip stress relaxation, promoting brittle or instability-driven fracture. At elevated temperature, thermally activated dislocation nucleation and glide can redistribute crack-tip stresses, promote local blunting, and delay catastrophic crack advance. The DBT in W is therefore not governed solely by bond breaking, but by the evolving balance between crack-tip stress concentration and localized plastic shielding \cite{gumbsch1998controlling,gumbsch2003,tarleton2009ddd,zhang2021tungsten}.

Alloying provides one route to modify this balance. Rhenium additions have historically been associated with improved ductility and reduced DBTT in W-based alloys, a phenomenon commonly referred to as the rhenium effect \cite{stephens1969,stephens1980,romaner2010effect,gornostyrev1991nature,caillard2020}. More recent studies, however, show that the Re effect is complex and depends on temperature, concentration, and microstructural state. Re can modify bonding, screw-dislocation core structure, and local lattice resistance, thereby altering the ease of plastic relaxation near defects and cracks. At the same time, Re solutes and other alloying additions can interact with dislocations and contribute to either softening or hardening, depending on chemistry, temperature, and microstructural state \cite{hu2017solute,zhang2024re,lin2025wre}. This broader picture is consistent with recent W-containing refractory alloy design studies showing that ductility is controlled by coupled composition, bonding, electronic-structure, and elastic-property descriptors rather than by a single alloying variable \cite{woodcox2025ductility,mahata2026ductility}. Understanding W--Re fracture behavior therefore requires mechanistic insight into how alloying alters crack-tip plasticity, not only empirical comparison of macroscopic ductility.

Atomistic simulation offers a direct route to interrogate these mechanisms. Molecular dynamics (MD) can resolve crack-tip stress redistribution, dislocation nucleation, crack-tip blunting, local structural disorder, and intermittent crack advance with atomic resolution \cite{armstrong2014,petersson2023molecular,alivaliollahi2023effect,lin2025wre}.
Recent atomic-scale studies of tungsten and W--Re systems have shown that fracture can involve a transition from cleavage-dominated crack growth to plasticity-assisted crack advance as temperature increases \cite{gumbsch2003,zhang2021tungsten,lin2025wre}. Nevertheless, MD simulations are restricted to nanometer length scales and strain rates many orders of magnitude higher than experimental testing. Transition temperatures inferred from MD should therefore be interpreted as high-rate mechanistic indicators rather than direct experimental DBTT predictions.

Atomistic simulation resolves the local processes by which dislocations nucleate, interact with the crack tip, and contribute to crack-tip blunting. However, it cannot access the larger microstructural length scales over which emitted dislocations expand, multiply, and interact collectively. A complete description of tungsten fracture therefore also requires a mesoscale view of dislocation-network evolution. Discrete dislocation dynamics (DDD) provides such a framework by resolving dislocation-source activation, line curvature, multiplication, and interactions with microstructural constraints \cite{bulatov2006,arsenlis2007,weygand2002}. Prior atomistic and DDD studies have shown how defect motion, precipitate strengthening, boundary-controlled pile-up formation, residual stress evolution, and hardening can be linked across scales \cite{MISTRY2025114122,mistry2026ppb}. DDD has also been applied to crack-tip shielding and ductile-to-brittle transition behavior in tungsten \cite{tarleton2009ddd}. These studies motivate the present use of DDD as a mesoscale extension of the crack-tip plasticity picture developed from the atomistic simulations.

In this study, atomistic fracture simulations are combined with a representative mesoscale dislocation dynamics calculation to examine temperature and composition dependent plasticity in W and W--Re systems. Molecular dynamics simulations of edge-cracked polycrystalline W are performed under displacement-controlled Mode-I tensile loading over a wide temperature range. The stress--strain response, instability stress and strain, pre-instability work density, and crack-tip defect structures are analyzed to characterize the competition between abrupt crack-tip instability and plastic relaxation. The DDD calculation separately examines how a prescribed source population evolves collectively within a polycrystalline tungsten aggregate. Together, the calculations provide complementary nanoscale and mesoscale evidence without treating the DDD calculation as a continuation of the MD trajectories or extrapolating the high-rate MD crossover directly to an experimental DBTT.

\section{Computational Methods}
\label{sec:methods}

\subsection{Molecular Dynamics Simulation}

\begin{figure}[htbp]
    \centering
    \includegraphics[width=0.50\textwidth]{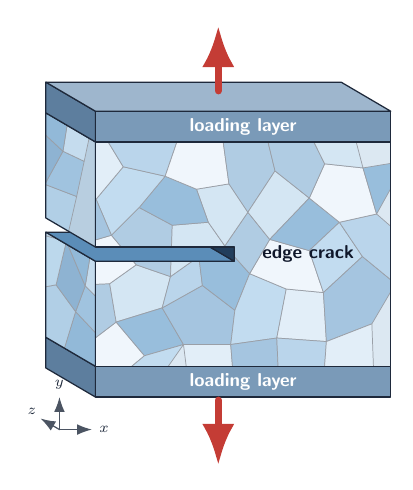}
    \caption{Schematic representation of the three-dimensional edge-cracked polycrystalline specimen subjected to displacement-controlled Mode-I tensile loading.}
    \label{fig:edge_crack_specimen}
\end{figure}

Molecular dynamics simulations were performed to investigate the effects of temperature and composition on fracture in edge-cracked polycrystalline W and W--Re under displacement-controlled Mode-I tensile loading. The initial polycrystalline simulation cell had dimensions of approximately $453\times402\times102$~\AA$^{3}$ ($45.3\times40.2\times10.2$~nm$^{3}$) and contained $\sim$1.2 million atoms. An atomistically sharp edge crack was introduced into the polycrystalline specimen to generate a localized crack-tip stress field during tensile deformation. Prior to relaxation, atoms separated by less than 0.5~\AA{} were removed to eliminate severe local overlaps, resulting in a final system containing $\sim$1.2 million atoms. The W--Re models were constructed using the same geometry and number of atomic sites, with Re introduced substitutionally at concentrations from 1 to 10~at.\% in increments of 1~at.\%. All simulations were carried out using the Large-scale Atomic/Molecular Massively Parallel Simulator (LAMMPS) \cite{thompson2022lammps}. Interatomic interactions were described using the embedded-atom method (EAM) potential developed by Bonny \textit{et al.} for plastic deformation in W--Re alloys \cite{Bonny2017WRe}. The same potential framework was used for both pure W and W--Re systems, thereby providing a consistent description of W--W, W--Re, and Re--Re interactions throughout the composition-dependent calculations. Initially, periodic boundary conditions were imposed in all three directions to facilitate structural relaxation. The system was relaxed using conjugate-gradient energy minimization together with isotropic box relaxation. Following minimization, free surfaces were introduced in the two in-plane directions using shrink-wrapped boundaries, while periodicity was retained along the out-of-plane crack-front direction (Fig.~\ref{fig:edge_crack_specimen}). The resulting boundary condition was therefore nonperiodic in the crack-opening plane and periodic along the crack front, allowing crack opening, surface formation, and local stress redistribution while retaining a quasi-three-dimensional crack-front geometry.

Following structural relaxation, each system was equilibrated at the target temperature in the canonical (NVT) ensemble. Initial atomic velocities were assigned from a Maxwell--Boltzmann distribution corresponding to the prescribed temperature, followed by 50,000 MD timesteps (50~ps) of equilibration using a Nose--Hoover thermostat with a damping parameter of 0.1~ps. A timestep of 1~fs was used throughout the calculations. Neighbor lists were constructed using a binning algorithm with a 2.0~\AA{} skin distance. A systematic temperature--composition simulation matrix was used to examine both the pure-W response and the influence of Re addition. Pure W was simulated from 300 to 1800~K at 100~K intervals, corresponding to 16 temperature conditions. The nine available W--Re compositions were evaluated over the same 16 temperatures, giving 144 W--Re simulations. The principal simulation parameters are summarized in Table~\ref{tab:md_parameters}.

\begin{table}[H]
    \centering
    \caption{Summary of the molecular dynamics fracture simulation matrix and principal computational parameters.}
    \label{tab:md_parameters}
    \begin{tabular}{ll}
        \hline
        \textbf{Parameter} & \textbf{Value} \\
        \hline
        Initial cell dimensions & $453\times402\times102$~\AA$^{3}$ \\
        Initial number of atoms & 1,191,915 \\
        Alloy system & W--Re \\
        Re concentration & 1--10~at.\% \\
        Temperature range & 300--1800~K \\
        Interatomic potential & EAM, Bonny \textit{et al.} \cite{Bonny2017WRe} \\
        MD timestep & 1~fs \\
        Equilibration time & 50~ps \\
        Applied strain rate & $5\times10^{8}$~s$^{-1}$ \\
        \hline
    \end{tabular}
\end{table}

After thermal equilibration, Mode-I loading was imposed through displacement controlled motion of two grip regions located at the upper and lower boundaries of the simulation cell. Equal and opposite velocities were prescribed to the two grips to produce symmetric crack opening, while atoms in the central crack-containing region remained dynamically mobile. The total relative grip velocity was determined from the instantaneous reference length in the loading direction to impose an engineering strain rate of $5\times10^{8}$~s$^{-1}$. The mobile atoms were thermostatted at the target temperature using the NVT ensemble, whereas the grip atoms were driven kinematically and were excluded from thermostat integration. This loading procedure introduced deformation through explicit boundary displacement rather than homogeneous affine deformation of the simulation box, allowing crack-tip blunting, dislocation emission, localized atomic rearrangement, intermittent crack advance, and stress relaxation to develop directly from the atomistic response. Each tensile-deformation calculation was propagated for up to 250,000 timesteps (250~ps). Atomic configurations for the full simulation cell were recorded every 10,000 timesteps, while a higher temporal resolution was employed for the crack-tip region. Atoms within the crack-tip region of interest were recorded every 500 timesteps, corresponding to a temporal resolution of 0.5~ps, for subsequent crack-tip structural and dislocation analysis. In addition, the global strain and tensile stress were recorded every 100 timesteps (0.1~ps) to provide the high-resolution mechanical response used in the stress--strain and instability analyses.

\subsection{Mechanical Metrics and Crossover Analysis}

The mechanical response was quantified from the global virial stress tensor over the analyzed interval \(\epsilon\leq0.05\). Engineering strain was calculated relative to the equilibrated loading-direction length immediately before tensile deformation. The instability stress, \(\sigma_{\mathrm{inst}}\), was defined as the global maximum of the processed tensile response within the analyzed strain window; this maximum precedes the principal load-relaxation event in the trajectories examined here. The corresponding strain was denoted \(\epsilon_{\mathrm{inst}}\). The work density to instability was calculated by trapezoidal integration,
\begin{equation}
    W_{\mathrm{inst}}=\int_0^{\epsilon_{\mathrm{inst}}}\sigma_{yy}(\epsilon)\,\mathrm{d}\epsilon.
    \label{eq:method_winst}
\end{equation}
Because stress is integrated with respect to dimensionless strain, \(W_{\mathrm{inst}}\) is a volumetric work density; it is not a fracture energy or critical energy-release rate. The effective initial tensile stiffness, \(E_{\mathrm{eff}}\), was obtained from a linear fit over \(0.001\leq\epsilon\leq0.006\) and characterizes the global response of the cracked specimen rather than an intrinsic elastic modulus. The same normalization and sigmoid-fitting procedure was applied independently to W--1Re through W--4Re. Fit quality was quantified using \(R^2\). Sensitivity of \(T_{\mathrm{MD}}^{*}\) to individual temperature
points was evaluated by repeating each fit after omitting one temperature.

For pure W and W--Re, the temperature-dependent work values were min--max normalized as
\begin{equation}
    \widehat{W}_{\mathrm{inst}}(T)=
    \frac{W_{\mathrm{inst}}(T)-W_{\min}}{W_{\max}-W_{\min}},
    \qquad
    D_E(T)=1-\widehat{W}_{\mathrm{inst}}(T).
    \label{eq:method_work_loss}
\end{equation}
The work-loss metric was fitted with
\begin{equation}
    D_E(T)=c_0+
    \frac{c_1}{1+\exp[-(T-T_{\mathrm{MD}}^{*})/\Delta T]}.
    \label{eq:method_sigmoid}
\end{equation}
The midpoint \(T_{\mathrm{MD}}^{*}\) and width \(\Delta T\) were selected by minimizing the sum of squared residuals on a grid spanning 500--1400~K in 0.5-K increments and 20--500~K in 1-K increments, respectively; \(c_0\) and \(c_1\) were solved by linear least squares for every grid pair. The fitted midpoint is reported to the precision supported by the 100-K simulation spacing and is used only as an internal high-rate crossover marker. To screen the available alloys, composition effects were evaluated at matched temperatures using \(\Delta W_{\mathrm{inst}}=W_{\mathrm{inst}}^{\mathrm{W-Re}}-W_{\mathrm{inst}}^{\mathrm{W}}\). W--10Re, which gives the largest mean positive change across the representative 700, 900, 1000, and 1200~K screening conditions, was then compared with pure W over all 16 temperatures. Atomistic configurations were analyzed using OVITO \cite{stukowski2010ovito}, and line defects were extracted using the Dislocation Extraction Algorithm (DXA) \cite{stukowski2010dxa}.

\subsection{Mesoscale Discrete Dislocation Dynamics Model}
\label{sec:ddd_method}
\begin{figure}[H]
    \centering
    \begin{subfigure}[b]{0.48\linewidth}
        \centering
        \includegraphics[width=\linewidth]{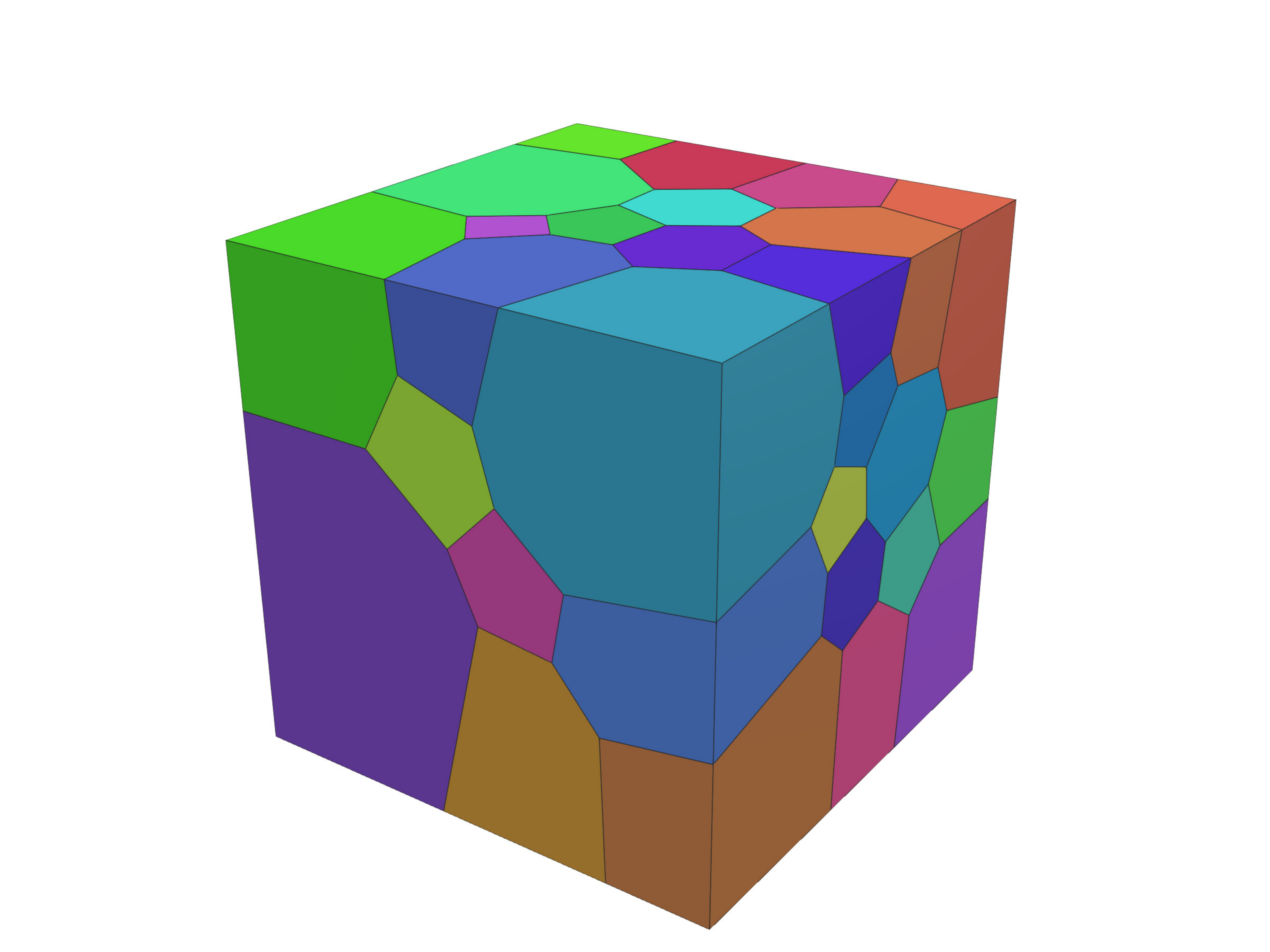}
        \caption{Opaque polycrystalline geometry.}
        \label{fig:ddd_poly_solid}
    \end{subfigure}
    \hfill
    \begin{subfigure}[b]{0.48\linewidth}
        \centering
        \includegraphics[width=\linewidth]{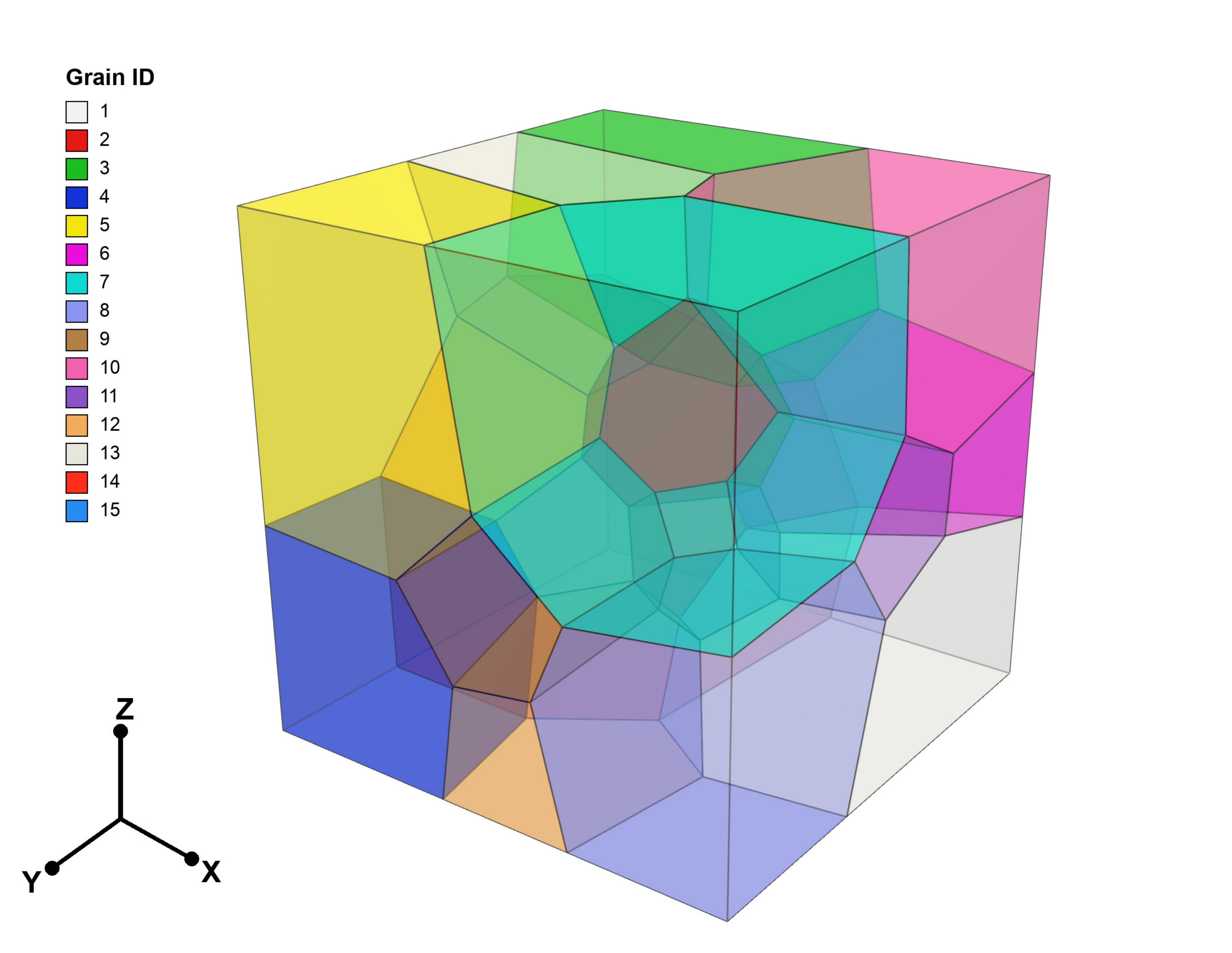}
        \caption{Transparent 15-grain geometry with Grain ID legend.}
        \label{fig:ddd_poly_transparent}
    \end{subfigure}
    \caption{Representative polycrystalline geometry used for the DDD calculation. 
    (a) Solid rendering of the three-dimensional grain structure. 
    (b) Transparent rendering of the 15-grain DDD domain showing the internal grain topology, categorical Grain ID color assignment, and global coordinate frame.}
    \label{fig:ddd_geometry}
\end{figure}

To complement the atomistic crack-tip simulations, one representative three-dimensional DDD calculation was performed for polycrystalline W. The calculation addresses source operation and collective network development at the micrometer scale. It contains no crack, Re chemistry, MD-derived mobility or nucleation parameters, or temperature sweep and is therefore neither a continuation of the atomistic trajectories nor a prediction of their crossover temperature. A synthetic \(4\times4\times4~\mu\mathrm{m}^{3}\) domain containing 15 grains was generated with Neper \cite{quey2011neper,quey2018neper}. The Neper facets in Fig.~\ref{fig:ddd_geometry} define the grain visualization and supply grain centroids, neighborhood relations, and orientations to the DDD setup. 

The four most highly stressed BCC \(\{110\}\langle111\rangle\) slip systems in each grain were selected using their Schmid factors under uniaxial loading along \(z\). This produced 60 prescribed Frank--Read-type source configurations with \(\frac{1}{2}\langle111\rangle\) Burgers vectors. These sources provide a controlled initial dislocation population and are not intended to reproduce the mixed defect structure obtained from the MD simulations.  Grain-boundary influence was represented by a heuristic global load correction rather than explicit segment transmission. At each recorded state, segment midpoints near a boundary were assigned to low- or high-angle categories using nearest-seed grain ownership and a 15\(^{\circ}\) misorientation threshold. The stress applied over the next block was
\begin{equation}
    \sigma_{\mathrm{eff}}=
    \max\!\left[\sigma_{\mathrm{app}}-f_{\mathrm{L}}(450~\mathrm{MPa})
    -0.97f_{\mathrm{H}}\sigma_{\mathrm{app}},\,0\right],
    \label{eq:ddd_effective_stress}
\end{equation}
where \(f_{\mathrm{L}}\) and \(f_{\mathrm{H}}\) are the fractions of segments identified near low- and high-angle boundaries. For the representative DDD calculation, a segment-count-based density proxy was evaluated as
\begin{equation}
    \rho_{\mathrm{seg}}(t)=\frac{N_{\mathrm{seg}}(t)\overline{L}_{\mathrm{seg}}}{V},
    \label{eq:ddd_density}
\end{equation}
where \(N_{\mathrm{seg}}\) is the number of discretized segments, \(\overline{L}_{\mathrm{seg}}=15b\) is an assumed mean segment length, and \(V\) is the polycrystal volume. Because dynamic remeshing changes the segment count, \(\rho_{\mathrm{seg}}\) is treated as a network-development proxy rather than an exact summed line-length density. A Taylor-type resistance indicator was post-processed as \(\tau_{\mathrm{T}}=\alpha\mu b\sqrt{\rho_{\mathrm{seg}}}\), using \(\alpha=0.3\), \(\mu=160.6~\mathrm{GPa}\), and \(b=2.741\times10^{-10}~\mathrm{m}\).

\section{Results and Discussion}
\label{sec:results}

\subsection{Atomistic Characteristics of Crack-Tip Deformation}
\label{sec:atomistic_crack_tip_results}

\begin{figure}[H]
    \centering
    \captionsetup{font=small}

    \begin{subfigure}[b]{0.48\linewidth}
        \centering
        \includegraphics[width=\linewidth]{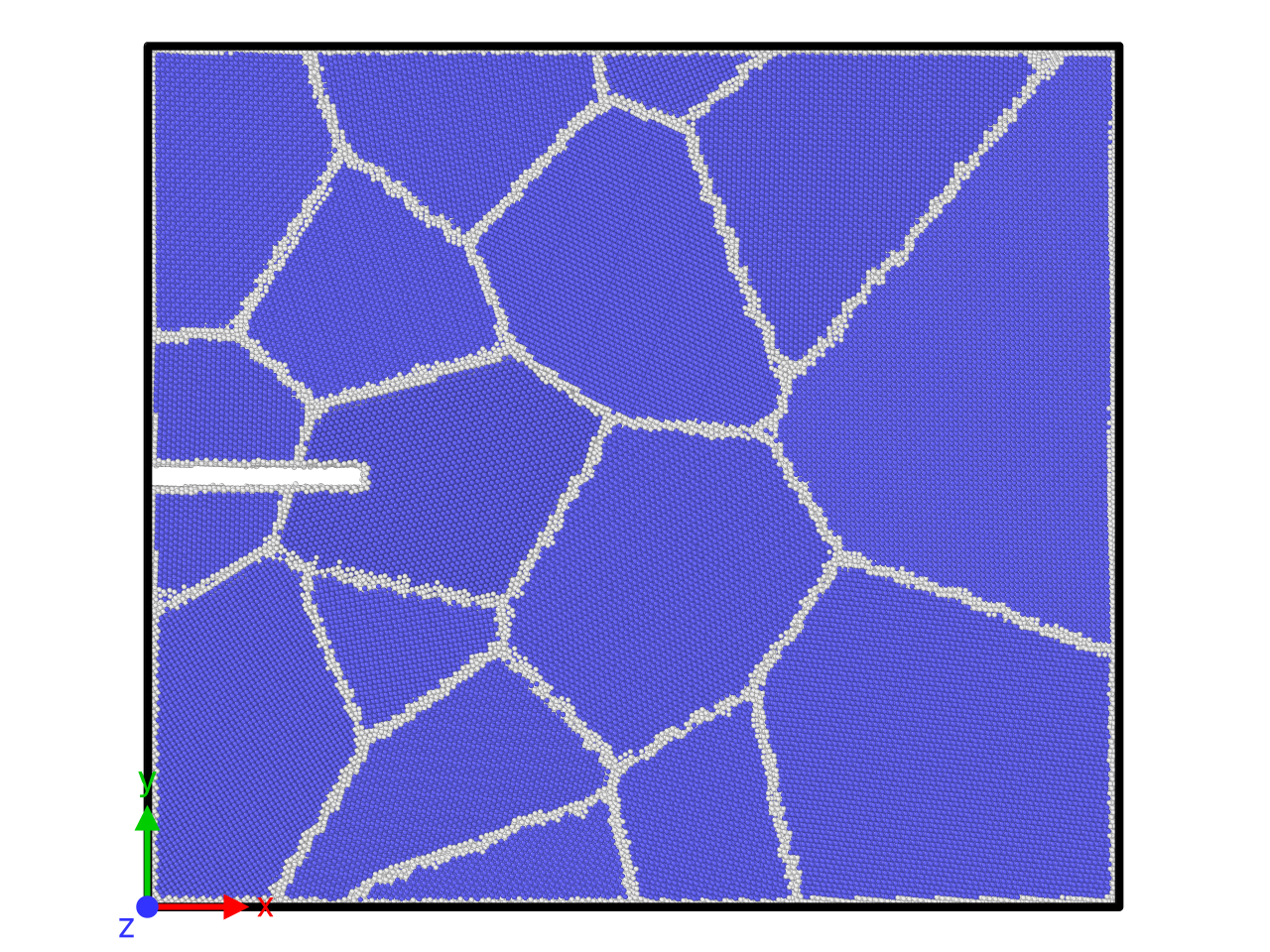}
        \par\smallskip\textbf{(a)}
    \end{subfigure}
    \hfill
    \begin{subfigure}[b]{0.48\linewidth}
        \centering
        \includegraphics[width=\linewidth]{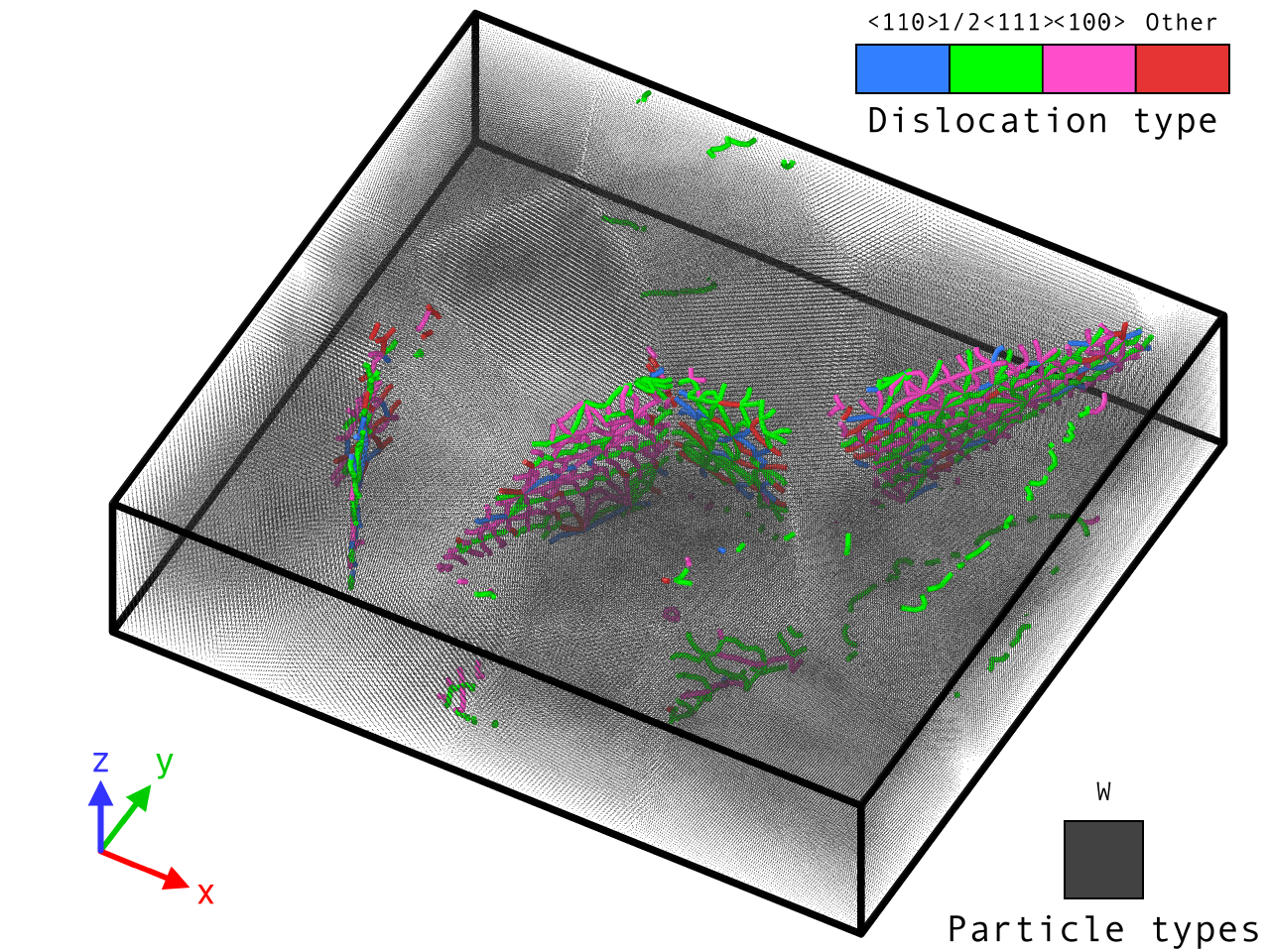}
        \par\smallskip\textbf{(b)}
    \end{subfigure}

    \vspace{0.20cm}

    \begin{subfigure}[b]{0.48\linewidth}
        \centering
        \includegraphics[width=\linewidth]{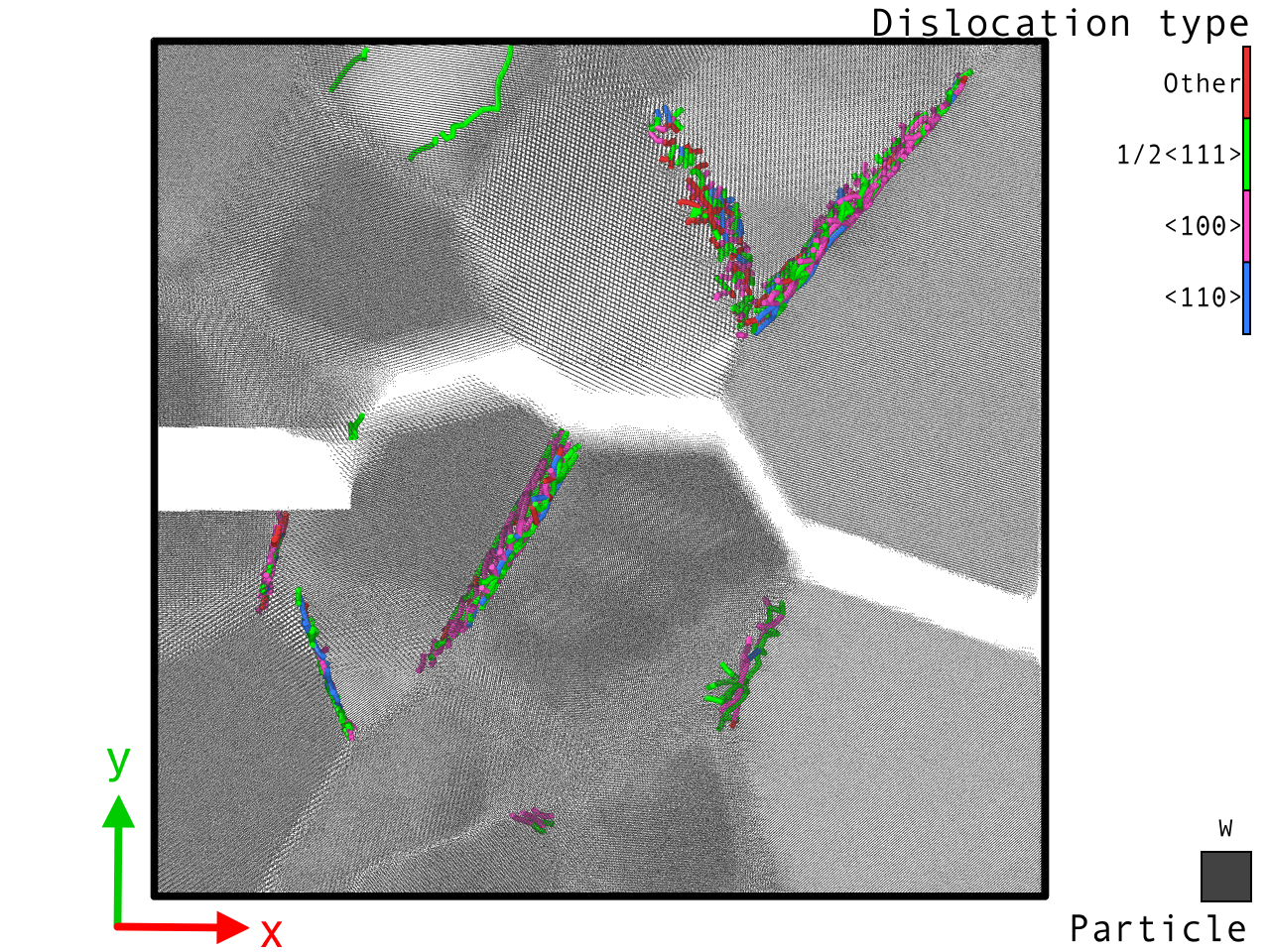}
        \par\smallskip\textbf{(c)}
    \end{subfigure}
    \hfill
    \begin{subfigure}[b]{0.48\linewidth}
        \centering
        \includegraphics[width=\linewidth]{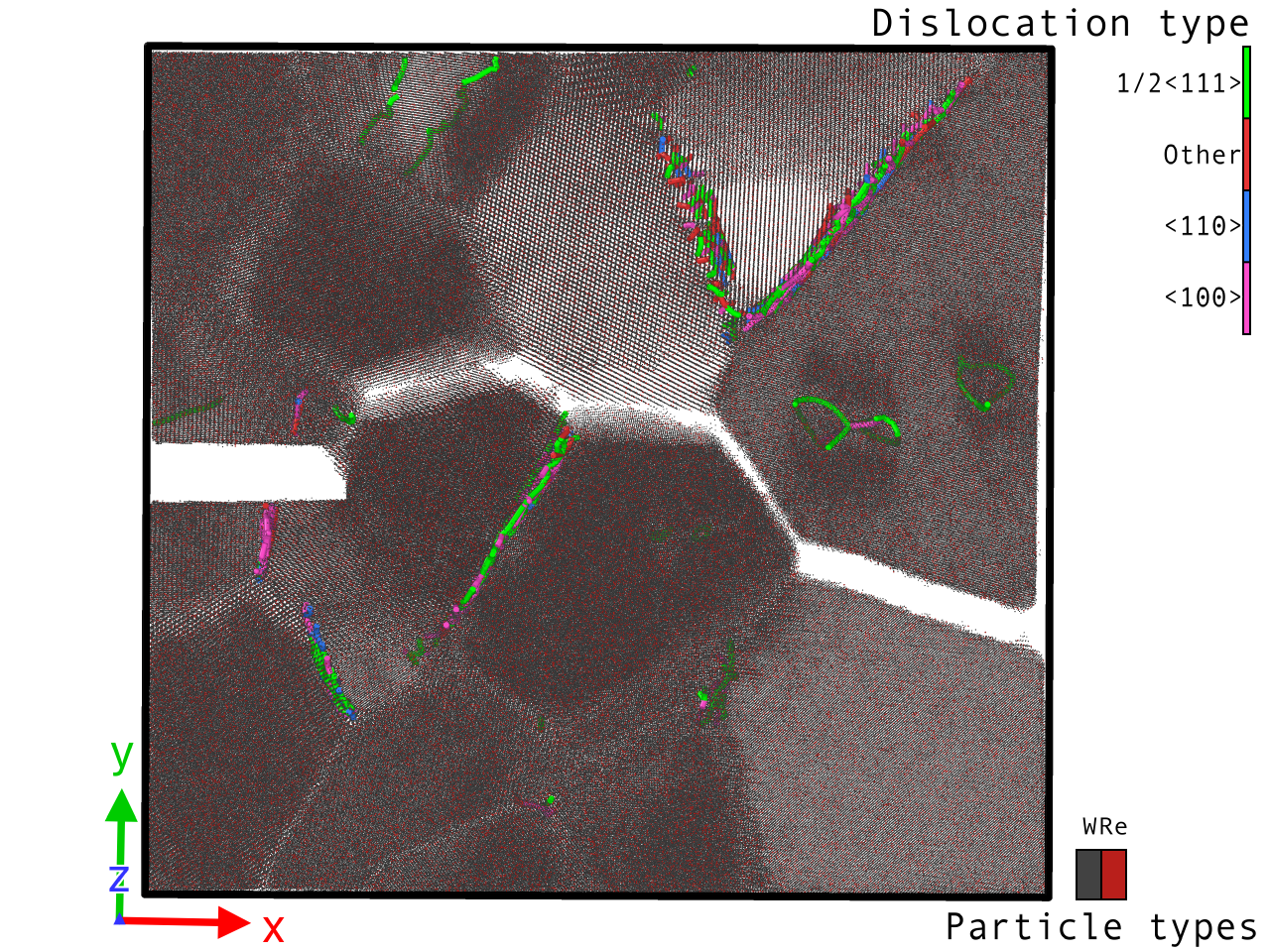}
        \par\smallskip\textbf{(d)}
    \end{subfigure}

    \caption{Representative atomistic configurations of edge-cracked polycrystalline W and W--10Re under Mode-I loading: 
    (a) undeformed reference, \(\epsilon=0\); 
    (b) pure W at 300~K, nominal \(\epsilon\approx0.005\); 
    (c) pure W at 300~K, nominal \(\epsilon\approx0.105\); and 
    (d) W--10Re at 1200~K, nominal \(\epsilon\approx0.075\). 
    In the DXA images, green, magenta, blue, and red lines denote 
    \(\frac{1}{2}\langle111\rangle\), \(\langle100\rangle\), 
    \(\langle110\rangle\), and other classifications, respectively. 
    The configurations illustrate representative deformation stages rather than matched temperature or strain states.}
    \label{fig:crack_tip_dxa_snapshots}
\end{figure}

Fig.~\ref{fig:crack_tip_dxa_snapshots} presents representative atomistic configurations from the initial edge-cracked polycrystal and from subsequent deformation of W and W--Re. The initially sharp crack intersects a heterogeneous grain structure, so the local crack-tip response is influenced by both the imposed Mode-I loading and the surrounding grain-boundary geometry. During loading, DXA-classified line segments develop in the crack-tip region and along nearby interfaces. Crack-tip deformation is therefore accompanied by localized defect activity and structural rearrangement rather than by geometrically ideal cleavage alone.

The extracted population is not restricted to the conventional \(\frac{1}{2}\langle111\rangle\) Burgers-vector family. The \(\langle100\rangle\)  \(\langle110\rangle\), and ``other'' DXA categories occur primarily in highly deformed regions near cracks, interfaces, and free surfaces. These segments can include transient reaction products, strongly curved line sections, and structures for which an unambiguous crystallographic assignment is difficult. Accordingly, the atomistic images establish a heterogeneous crack-tip defect field, while the edge/screw analysis presented below is restricted to the conventional \(\frac{1}{2}\langle111\rangle\) population.

\subsection{Temperature-Dependent Mechanical Instability in Pure W}
\label{sec:pure_w_mechanical_results}

\begin{figure}[H]
    \centering
    \includegraphics[width=0.80\linewidth]{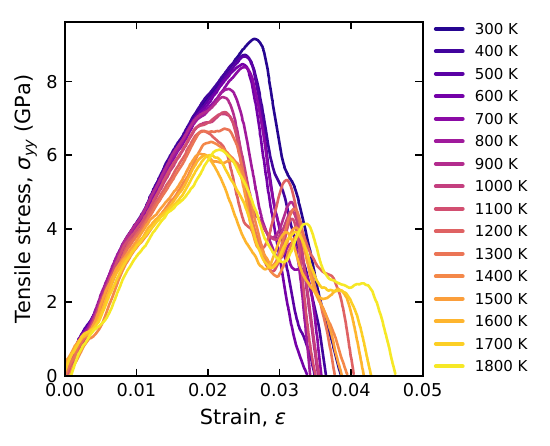}
    \caption{Temperature-dependent tensile response of edge-cracked polycrystalline W under displacement-controlled Mode-I loading.}
    \label{fig:w_stress_strain}
\end{figure}

The mechanical response of the edge-cracked pure-W specimen was evaluated from 300 to 1800~K. Fig.~\ref{fig:w_stress_strain} shows the baseline-corrected tensile stress--strain curves. All curves exhibit an initial loading regime followed by a maximum stress and a major stress-relaxation event. The global maximum of the processed tensile response within \(\epsilon\leq0.05\) is used consistently as the operational onset of mechanical instability in the present simulations.

The instability stress decreases overall with increasing temperature, demonstrating thermal weakening of the cracked specimen under the imposed high-rate loading. The low-temperature trajectories generally undergo a comparatively sharp loss of load-carrying capacity after the operational stress maximum. At higher temperature, the post-maximum response contains more pronounced shoulders and secondary relaxation events. These features demonstrate repeated load redistribution after the operational maximum; the stress signal alone does not uniquely separate crack advance, inertial oscillation, and local plastic rearrangement. The atomistic configurations in Fig.~\ref{fig:crack_tip_dxa_snapshots} nevertheless show that defect activity accompanies this post-maximum evolution.
To quantify the mechanical state reached at the operational instability, the work density to instability was calculated as
\begin{equation}
W_{\mathrm{inst}}
=
\int_{0}^{\epsilon_{\mathrm{inst}}}
\sigma_{yy}(\epsilon)\,\mathrm{d}\epsilon,
\label{eq:winst}
\end{equation}
where \(\epsilon_{\mathrm{inst}}\) is the strain at the baseline-corrected stress maximum. Since stress is integrated with respect to strain, \(W_{\mathrm{inst}}\) is a volumetric mechanical work density. It is not a critical energy-release rate or a fracture energy per newly created crack area.

\begin{figure}[htbp]
    \centering
    \includegraphics[width=0.85\linewidth]{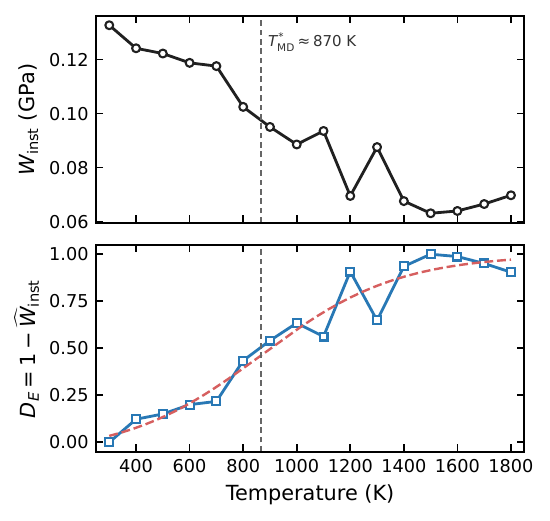}
        \caption{Energy-based characterization of the high-rate transition regime in edge-cracked polycrystalline W. The upper panel shows the work density to instability, \(W_{\mathrm{inst}}\), evaluated from the baseline-corrected stress up to its global maximum within \(\epsilon\leq0.05\). The lower panel shows the normalized work-loss metric, \(D_E=1-\widehat{W}_{\mathrm{inst}}\). The vertical line marks the midpoint of the fitted sigmoid, reported as \(T_{\mathrm{MD}}^{*}\approx870~\mathrm{K}\).}
    \label{fig:w_energy_dbt}
\end{figure}

Fig.~\ref{fig:w_energy_dbt} shows an overall decrease in \(W_{\mathrm{inst}}\) with increasing temperature, together with local non-monotonic fluctuations. The reduction results from the combined changes in the stress level and the strain at the operational stress maximum. A normalized work-loss metric was defined as
\begin{equation}
D_E=1-\widehat{W}_{\mathrm{inst}},
\label{eq:energy_loss_metric}
\end{equation}
where \(\widehat{W}_{\mathrm{inst}}\) denotes min--max normalization over the simulated temperature range. A sigmoid fitted to \(D_E\) gives a midpoint of 868~K and a width parameter of 255~K. Because the simulations are spaced at 100-K intervals and the fitted crossover is broad, the midpoint is reported as \(T_{\mathrm{MD}}^{*}\approx870~\mathrm{K}\). This marker is used below to distinguish conditions below, near, and above the energy-based high-rate crossover. The energy metric does not, by itself, establish a unique fracture-mode transition.

\begin{figure}[htbp]
    \centering
    \includegraphics[width=0.85\linewidth]{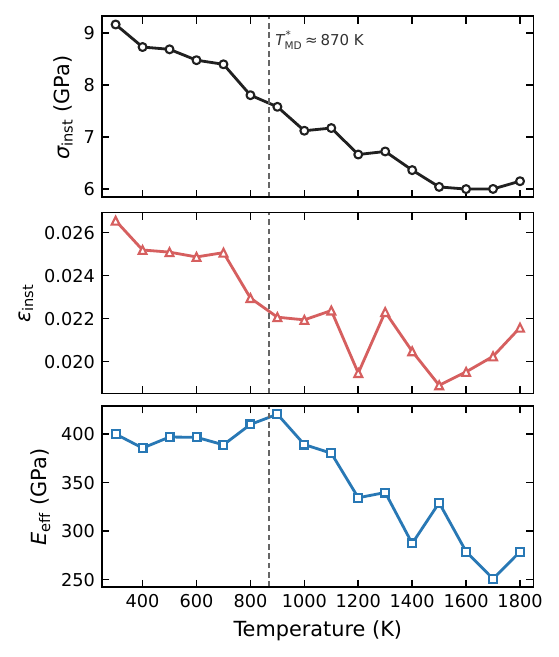}
    \caption{Temperature dependence of the baseline-corrected instability stress, \(\sigma_{\mathrm{inst}}\), instability strain, \(\epsilon_{\mathrm{inst}}\), and effective initial tensile stiffness, \(E_{\mathrm{eff}}\), of the edge-cracked pure-W specimen. }
    \label{fig:w_strength_metrics}
\end{figure}

The extracted metrics in Fig.~\ref{fig:w_strength_metrics} separate three aspects of the response. First, \(\sigma_{\mathrm{inst}}\) decreases strongly over the simulated temperature range. Second, \(E_{\mathrm{eff}}\) also decreases overall, although the variation is not monotonic at every temperature. Because this stiffness is extracted from a cracked polycrystalline body with free surfaces and grip loading, it represents the initial compliance of the simulated specimen rather than only lattice elasticity. Third, \(\epsilon_{\mathrm{inst}}\) is more scattered than either \(\sigma_{\mathrm{inst}}\) or \(W_{\mathrm{inst}}\). Heating therefore reduces the stress and work density associated with the operational instability more clearly than it shifts that instability to a larger strain.

\subsection{Temperature-Dependent Dislocation Character in Pure W}
\label{sec:pure_w_dislocation_results}

The defect-character comparison was performed at a common engineering strain of \(\epsilon=0.045\), which lies after the operational global processed-stress maximum for the conditions considered. Restricting the comparison to a common strain avoids conflating temperature-dependent trajectory lengths with changes in dislocation density. Only DXA-classified \(\frac{1}{2}\langle111\rangle\) segments are separated into edge and screw components, and the instantaneous simulation-cell volume is used in the density calculation.

\begin{figure}[htbp]
    \centering
    \includegraphics[width=\linewidth]{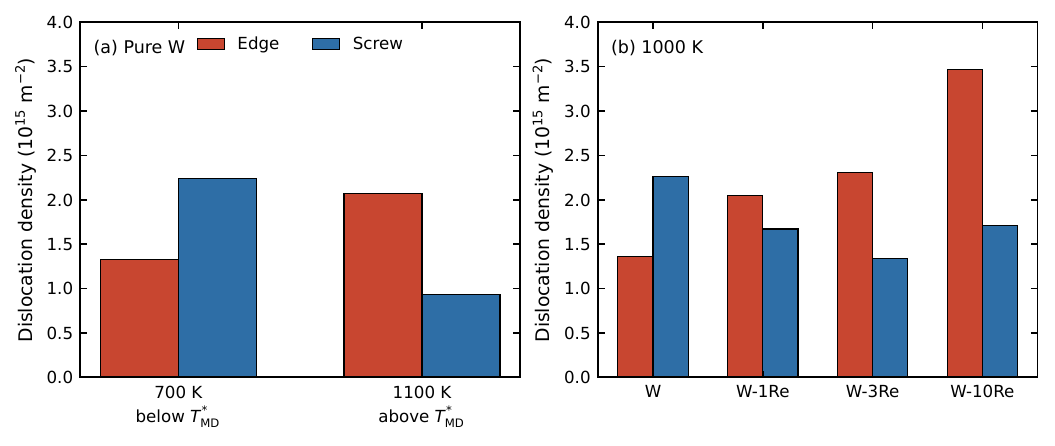}
    \caption{DXA-derived \(\frac{1}{2}\langle111\rangle\) edge and screw dislocation densities at the common engineering strain \(\epsilon=0.045\). (a) Pure W at two representative conditions on the lower- and higher-temperature sides of the energy-based crossover: 700 and 1100~K. (b) Composition dependence at 1000~K for W, W--1Re, W--3Re, and W--10Re. The comparison characterizes the retained post-instability defect population at a common state and does not directly measure dislocation mobility.}
    \label{fig:md_common_strain_edge_screw}
\end{figure}

For pure W at 700~K, the edge and screw densities are approximately \(1.33\times10^{15}\) and \(2.24\times10^{15}~\mathrm{m}^{-2}\), respectively, giving an edge fraction of 0.37. At 1100~K, the edge density is approximately \(2.07\times10^{15}~\mathrm{m}^{-2}\), whereas the screw density is approximately \(0.94\times10^{15}~\mathrm{m}^{-2}\), giving an edge fraction of 0.69. Between these two representative conditions, the retained post-instability population changes from screw-dominated to edge-dominated. This two-temperature comparison establishes a difference between the selected states, not a continuous temperature-dependent regime boundary. It also does not imply that the total dislocation density must increase with temperature. The mechanical and defect data together instead indicate that the selected higher-temperature state differs more clearly in retained line character and load-relaxation mode than in a monotonic increase of peak dislocation density. Additional temperature-resolved dislocation histories for pure W are provided in Supplementary Figs.~\ref{supp-fig:supp_edge_screw_same_panel}--\ref{supp-fig:supp_w_timelines}. 
The central atomistic result is that temperature changes both the mechanical threshold for instability and the character of the defect population retained during crack-tip deformation. The instability stress, effective specimen stiffness, and work density accumulated up to the global processed-stress maximum within \(\epsilon\leq0.05\) decrease overall with increasing temperature. These trends establish thermal weakening under the present loading protocol, but they do not imply that the material becomes more brittle with temperature. In BCC W, the ductile-to-brittle transition is governed by a kinetic competition: crack advance is favored when the crack-tip stress intensification develops faster than dislocations can nucleate and redistribute stress, whereas plastic shielding becomes more effective when defect nucleation and motion can compete with crack propagation \cite{gumbsch1998controlling,gumbsch2003,giannattasio2007brittle,giannattasio2011loading}.

The present results are consistent with this competition but reveal it through several complementary observables. The stress response changes from comparatively abrupt post-maximum load loss at selected lower temperatures to more distributed relaxation at selected higher temperatures. At the common strain \(\epsilon=0.045\), the retained \(\frac{1}{2}\langle111\rangle\) population is screw-dominated at the representative 700-K condition and edge-dominated at 1100~K. These two states do not by themselves establish a continuous line-character transition across the full temperature range. They do show that the relevant distinction is not a monotonic increase in total dislocation density. A lower retained density can coexist with effective relaxation if dislocations are transported away from the most highly stressed region, annihilate, react, or change line character. Conversely, a high local density can indicate restricted transport and accumulation rather than efficient shielding. The edge/screw partition and post-maximum stress morphology therefore provide more defensible mechanistic evidence than peak density alone.

The mixed DXA population further indicates that the crack-tip region cannot be represented as an isolated ideal source on a single slip system. The conventional \(\frac{1}{2}\langle111\rangle\) family remains the physically interpretable BCC slip population, while the non-\(\frac{1}{2}\langle111\rangle\) categories reflect the severe structural complexity near cracks, interfaces, and free surfaces. Similar atomistic and experimental studies have associated the elevated-temperature response of W with enhanced dislocation-mediated crack-tip accommodation and non-cleavage deformation \cite{zhang2021tungsten,gumbsch2003,schwiedrzik2018micro}. The present simulations support that broader picture, while showing that a polycrystalline crack-tip field produces substantial spatial and crystallographic heterogeneity.

The fitted midpoint \(T_{\mathrm{MD}}^{*}\approx870~\mathrm{K}\) should be interpreted as a characteristic crossover of the selected work-loss metric under the present atomistic loading protocol. It is not a universal material constant. First, \(W_{\mathrm{inst}}\) is dominated by the stress and strain accumulated up to the global processed-stress maximum within \(\epsilon\leq0.05\) and therefore contains a strong thermal-softening contribution. Second, the underlying temperature series is locally non-monotonic and the fitted width is approximately 255~K, so the midpoint represents a broad crossover rather than an abrupt boundary. Third, the fitted value depends on applying the stated initial-stress correction, smoothing rule, and instability criterion consistently. Finally, MD strain rates are many orders of magnitude higher than laboratory fracture-testing rates. The pronounced rate dependence of the tungsten transition is well established experimentally and mechanistically \cite{giannattasio2007brittle,giannattasio2011loading,gumbsch1998controlling_dbtt}.

The energy-based marker is nevertheless useful when it is treated as an internal reference and corroborated by independent structural information. In the present data, temperatures on opposite sides of \(T_{\mathrm{MD}}^{*}\) show different post-instability relaxation patterns and a reversal of the edge/screw balance at a common strain. The transition interpretation therefore rests on the conjunction of mechanical and defect-character evidence, not on the sigmoid fit alone. Reporting the value as an MD-derived high-rate crossover also avoids direct extrapolation to macroscopic DBTT measurements, which depend on loading rate, crack geometry, grain structure, impurity content, processing history, and specimen scale.

\subsection{Composition Dependence and the W--10Re Response}
\label{sec:wre_results}

The available W--Re compositions were screened at 700, 900, 1000, and 1200~K using the difference in work density to instability relative to pure W,
\begin{equation}
\Delta W_{\mathrm{inst}}
=
W_{\mathrm{inst}}^{\mathrm{W-Re}}
-
W_{\mathrm{inst}}^{\mathrm{W}}.
\label{eq:delta_winst}
\end{equation}
Positive values therefore indicate a larger area under the baseline-corrected stress--strain curve up to the global processed-stress maximum within \(\epsilon\leq0.05\), not a directly measured fracture toughness.

\begin{figure}[htbp]
    \centering
    \includegraphics[width=0.85\linewidth]{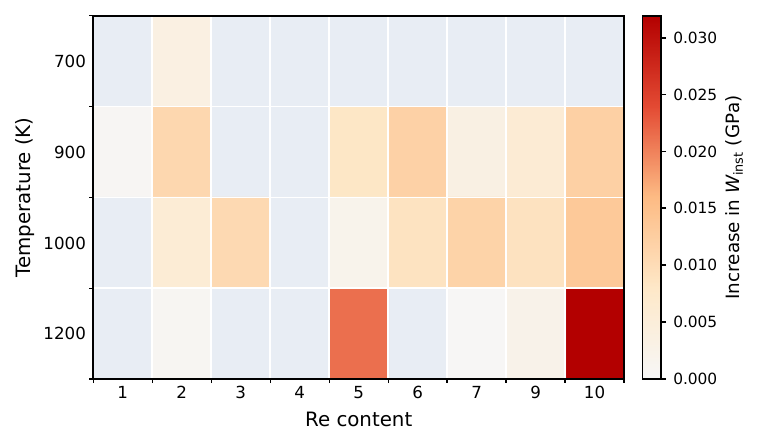}
    \caption{Screening of the available W--Re compositions using \(\Delta W_{\mathrm{inst}}\) relative to pure W at the same temperature. Positive values denote an increase in pre-instability work density, whereas values below zero denote a reduction. W--10Re gives the largest mean positive change among the available compositions and is therefore selected for the detailed comparison. The selection is specific to the simulated compositions, temperatures, microstructure, potential, and loading protocol.}
    \label{fig:wre_composition_screening}
\end{figure}

The composition map in Fig.~\ref{fig:wre_composition_screening} is non-monotonic, showing that Re content alone does not determine the response. Some dilute and intermediate compositions produce either small or temperature-dependent changes relative to pure W. Among the available datasets, W--10Re gives the largest mean increase in \(W_{\mathrm{inst}}\), with the clearest positive differences at 900--1200~K. W--10Re is consequently used as a representative Re-containing system rather than being identified as a universal optimum composition.

\begin{figure}[H]
    \centering
    \includegraphics[width=\linewidth]
    {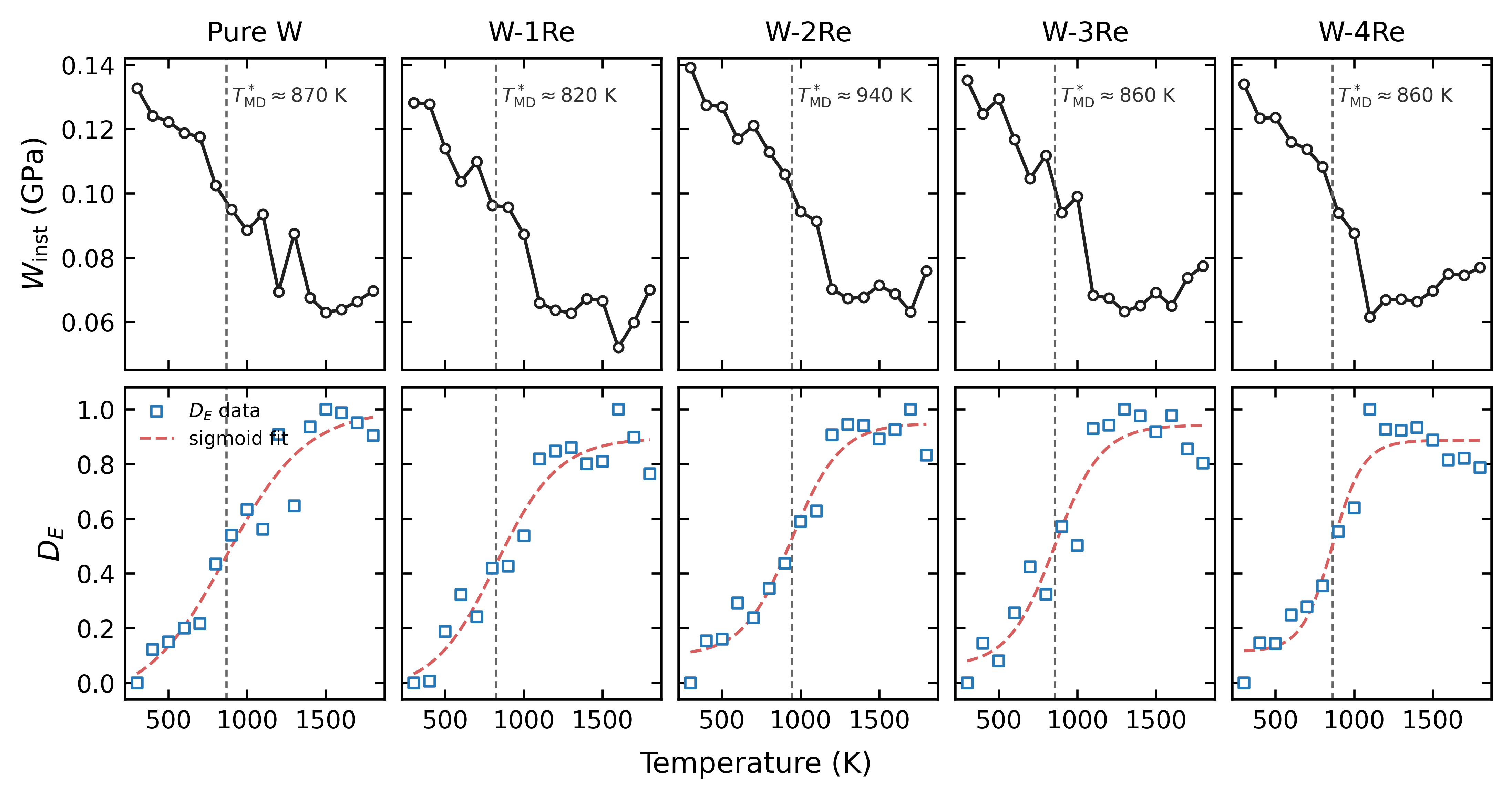}
    \caption{Composition-resolved work-based crossover analysis for pure W
    and W--1Re through W--4Re. The upper row shows the work density to
    instability, \(W_{\mathrm{inst}}\). The lower row shows the normalized
    work-loss data, \(D_E\), and the corresponding sigmoid fits. Vertical
    lines denote the fitted midpoints \(T_{\mathrm{MD}}^{*}\).}
    \label{fig:dilute_re_crossover_fits}
\end{figure}

\begin{figure}[H]
    \centering
    \includegraphics[width=0.5\linewidth]
    {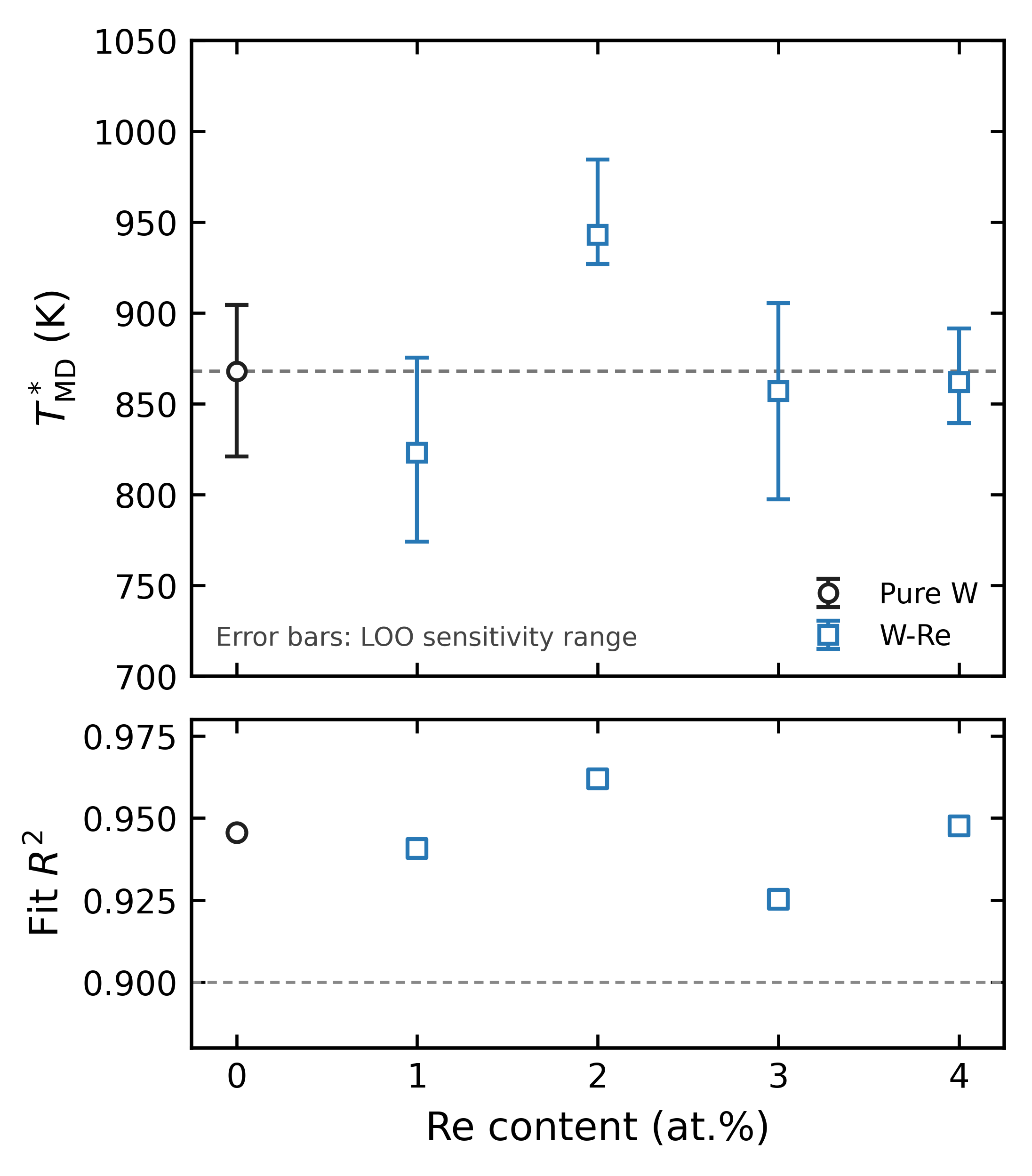}
    \caption{Composition dependence of the fitted work-based crossover for
    pure W and W--1Re through W--4Re. The upper panel shows
    \(T_{\mathrm{MD}}^{*}\), with error bars representing
    leave-one-temperature-out sensitivity ranges. The lower panel reports
    the sigmoid-fit \(R^2\).}
    \label{fig:dilute_re_crossover_summary}
\end{figure}
To determine whether dilute Re produces a systematic shift of the work-based crossover, the same analysis was applied to W--1Re through W--4Re. The fitted midpoints span 823--943~K, compared with 868~K for pure W, and all fits give \(R^2>0.92\) (Figs.~\ref{fig:dilute_re_crossover_fits} and
\ref{fig:dilute_re_crossover_summary}). Although W--2Re gives the highest midpoint, the composition series is non-monotonic, while W--1Re, W--3Re, and W--4Re remain comparatively close to pure W. Dilute Re
addition therefore does not produce a systematic shift of the operational high-rate crossover over 1--4~at.\% under the present protocol.

\begin{figure}[H]
    \centering
    \includegraphics[width=0.85\linewidth]{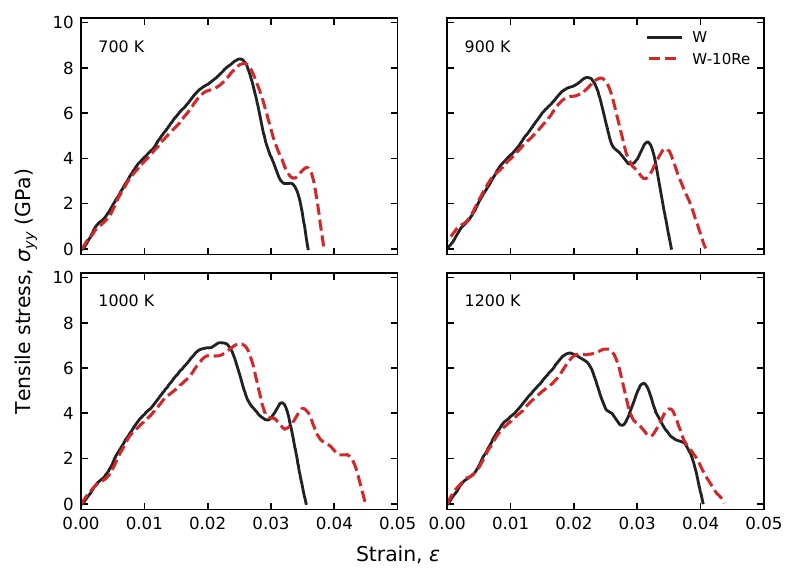}
    \caption{Baseline-corrected tensile stress--strain response of pure W and W--10Re at 700, 900, 1000, and 1200~K. The curves are shown through the principal post-instability load-relaxation event. Re addition changes the strain and work density associated with instability more consistently than it changes the maximum stress.}
    \label{fig:w_w10re_stress_strain}
\end{figure}

Fig.~\ref{fig:w_w10re_stress_strain} demonstrates that the Re effect cannot be described as uniform strengthening. At 700~K, W--10Re has a slightly lower instability stress and work density than pure W, while the instability strain changes only marginally. At 900 and 1000~K, the two systems reach similar global processed-stress maxima, but W--10Re reaches that maximum at a larger strain and accumulates more work beforehand. At 1200~K, the separation is largest among the four representative conditions: the increase in instability strain is approximately 0.0055 and the increase in \(W_{\mathrm{inst}}\) is approximately 0.032~GPa. The improvement at these temperatures therefore arises primarily from extending the pre-instability deformation interval rather than from raising the maximum tensile stress.

\begin{figure}[H]
    \centering
    \includegraphics[width=0.80\linewidth]{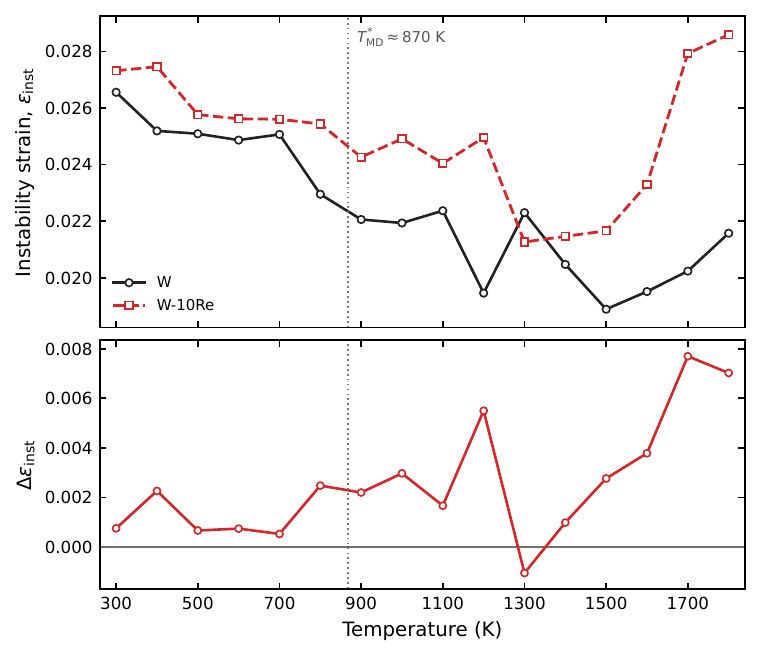}
    \caption{Temperature dependence of the instability strain in pure W and W--10Re. The upper panel shows \(\epsilon_{\mathrm{inst}}\), and the lower panel shows \(\Delta\epsilon_{\mathrm{inst}}=\epsilon_{\mathrm{inst}}^{\mathrm{W-10Re}}-\epsilon_{\mathrm{inst}}^{\mathrm{W}}\). The vertical line denotes \(T_{\mathrm{MD}}^{*}\). Positive values indicate that W--10Re reaches the global maximum of the processed stress within \(\epsilon\leq0.05\) at a larger strain than pure W.}
    \label{fig:w_w10re_instability_strain}
\end{figure}

The full temperature comparison in Fig.~\ref{fig:w_w10re_instability_strain} shows a positive \(\Delta\epsilon_{\mathrm{inst}}\) at 15 of the 16 simulated temperatures, with one negative excursion at 1300~K. The difference is generally modest below the crossover and becomes larger at several elevated-temperature conditions, particularly from 1500 to 1800~K. The peak-stress difference changes sign across the same range, confirming that the principal W--10Re response is a shift in the deformation sustained before instability rather than a composition-independent increase in strength.

The composition dependence of the \(\frac{1}{2}\langle111\rangle\) population at 1000~K is shown in Fig.~\ref{fig:md_common_strain_edge_screw}(b). At \(\epsilon=0.045\), the edge density increases from approximately \(1.36\times10^{15}~\mathrm{m}^{-2}\) in W to \(2.05\times10^{15}\), \(2.31\times10^{15}\), and \(3.47\times10^{15}~\mathrm{m}^{-2}\) in W--1Re, W--3Re, and W--10Re, respectively. The corresponding edge fraction rises from 0.38 in W to 0.55, 0.63, and 0.67. The screw density does not vary monotonically with Re content. The composition effect is therefore expressed primarily as a redistribution toward a larger retained edge-character population, accompanied by a particularly large total \(\frac{1}{2}\langle111\rangle\) density in W--10Re. Additional composition- and temperature-resolved edge/screw comparisons are provided in Supplementary Figs.~\ref{supp-fig:supp_burgers_density_300K}--\ref{supp-fig:supp_screw_density_w_wre} and \ref{supp-fig:supp_density_re_all_temperatures}--\ref{supp-fig:supp_temperature_effect_w_wre}.

\begin{figure}[H]
    \centering
    \includegraphics[width=0.85\linewidth]{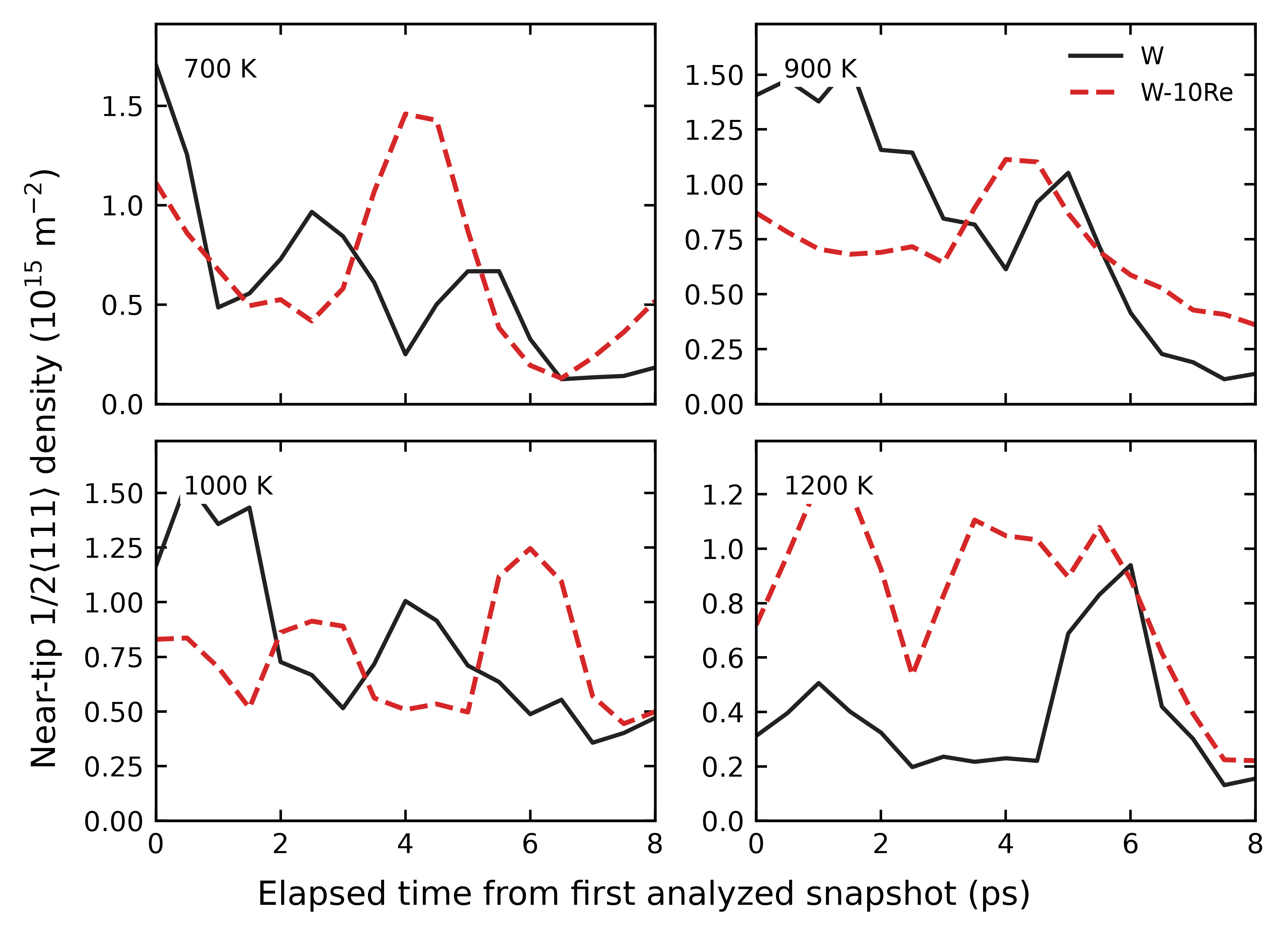}
    \caption{Evolution of the near-tip \(\frac{1}{2}\langle111\rangle\) density in pure W and W--10Re over the common 0--8~ps analysis interval at representative temperatures.}
    \label{fig:w_w10re_dislocation_activity}
\end{figure}

The common-window near-tip trajectories in Fig.~\ref{fig:w_w10re_dislocation_activity} reveal a strongly temperature-dependent Re effect. At 700, 900, and 1000~K, the maximum near-tip \(\frac{1}{2}\langle111\rangle\) dislocation density is lower in W--10Re than in pure W. The corresponding time-averaged difference is small and positive at 700~K (approximately \(0.05 \times 10^{15}\)~m\(^{-2}\)) but negative at 900 and 1000~K (approximately \(0.13\) and \(0.06 \times 10^{15}\)~m\(^{-2}\), respectively). At 1200~K, W--10Re exhibits both a higher peak and a larger time-averaged dislocation density, exceeding pure W by approximately \(0.12\) and \(0.45 \times 10^{15}\)~m\(^{-2}\), respectively. Re therefore does not uniformly increase the amount or persistence of near-tip dislocations. Instead, it modifies the magnitude and temporal distribution of the local \(\frac{1}{2}\langle111\rangle\) population, with the clearest enhancement occurring at 1200~K. The corresponding time-resolved W--10Re defect indicators are provided in Supplementary Fig.~\ref{supp-fig:supp_wre_timelines}. The W--Re results show that Re does not produce a single composition-independent strengthening or softening response. The instability-stress difference changes sign with temperature, and the composition screening is non-monotonic. In contrast, W--10Re reaches the operational processed-stress maximum at a larger strain than pure W at nearly every simulated temperature and exhibits a larger pre-instability work density at most conditions. Near 900--1200~K, the stress maxima of W and W--10Re remain comparable while the alloy sustains a longer loading interval. The dominant Re effect in this regime is therefore delayed mechanical instability rather than increased peak strength.

The defect analysis provides a compatible, though not independently causal, explanation. At 1000~K and \(\epsilon=0.045\), increasing Re content produces a monotonic increase in the edge fraction of the retained \(\frac{1}{2}\langle111\rangle\) population, while the screw density remains non-monotonic. Over the common 0--8~ps interval, W--10Re has a smaller maximum near-tip line length at 700, 900, and 1000~K; its time-averaged value is similar to pure W at 700~K and lower at 900 and 1000~K. Only at 1200~K are both the peak and time-averaged values clearly higher in the alloy. The delayed global instability of W--10Re therefore does not require a uniformly larger local line population. Together with the edge-character enrichment at a common strain, this result is consistent with Re modifying the character, timing, and spatial organization of crack-tip accommodation rather than merely increasing the number of dislocations.

This interpretation agrees qualitatively with the established complexity of the rhenium effect. Prior studies have shown that Re can alter screw-dislocation core structure, lattice resistance, and the balance between solid-solution softening and hardening in W \cite{romaner2010effect,li2012dislocation,hu2017solute,caillard2020,zhang2024re}. Experimental and atomistic investigations also demonstrate that the macroscopic effect depends on concentration, temperature, and microstructure \cite{stephens1969,wurster2010wrefracture,lin2025wre}. The present calculations do not directly determine a Peierls barrier, kink-pair activation energy, or mobility law. Consequently, the observed edge-character enrichment and delayed instability should be described as being consistent with altered defect-mediated relaxation, not as direct proof of a specific core-level mechanism.

The finding that W--10Re performs best among the available screening cases must also remain conditional. Only one polycrystalline geometry and one solute realization are represented at each temperature--composition point, and the composition matrix excludes an unavailable concentration. The result identifies W--10Re as the strongest case within this dataset and loading protocol; it does not establish a universal optimum Re content for fracture resistance.
\subsection{Representative Mesoscale Dislocation-Network Evolution}
\label{sec:ddd_results}

The DDD calculation provides a separate mesoscale demonstration of source-like bow-out and network development in polycrystalline W. The simulated \(4\times4\times4~\mu\mathrm{m}^{3}\) domain contains 15 grains and 60 initially prescribed Frank--Read-type source configurations and is subjected to a uniaxial applied stress of 1200~MPa. This calculation does not reproduce the MD crack geometry, temperature sweep, Re chemistry, or atomistically fitted mobility parameters.

\begin{figure}[htbp]
    \centering
    \begin{subfigure}[b]{0.48\textwidth}
        \centering
        \includegraphics[width=0.80\textwidth]{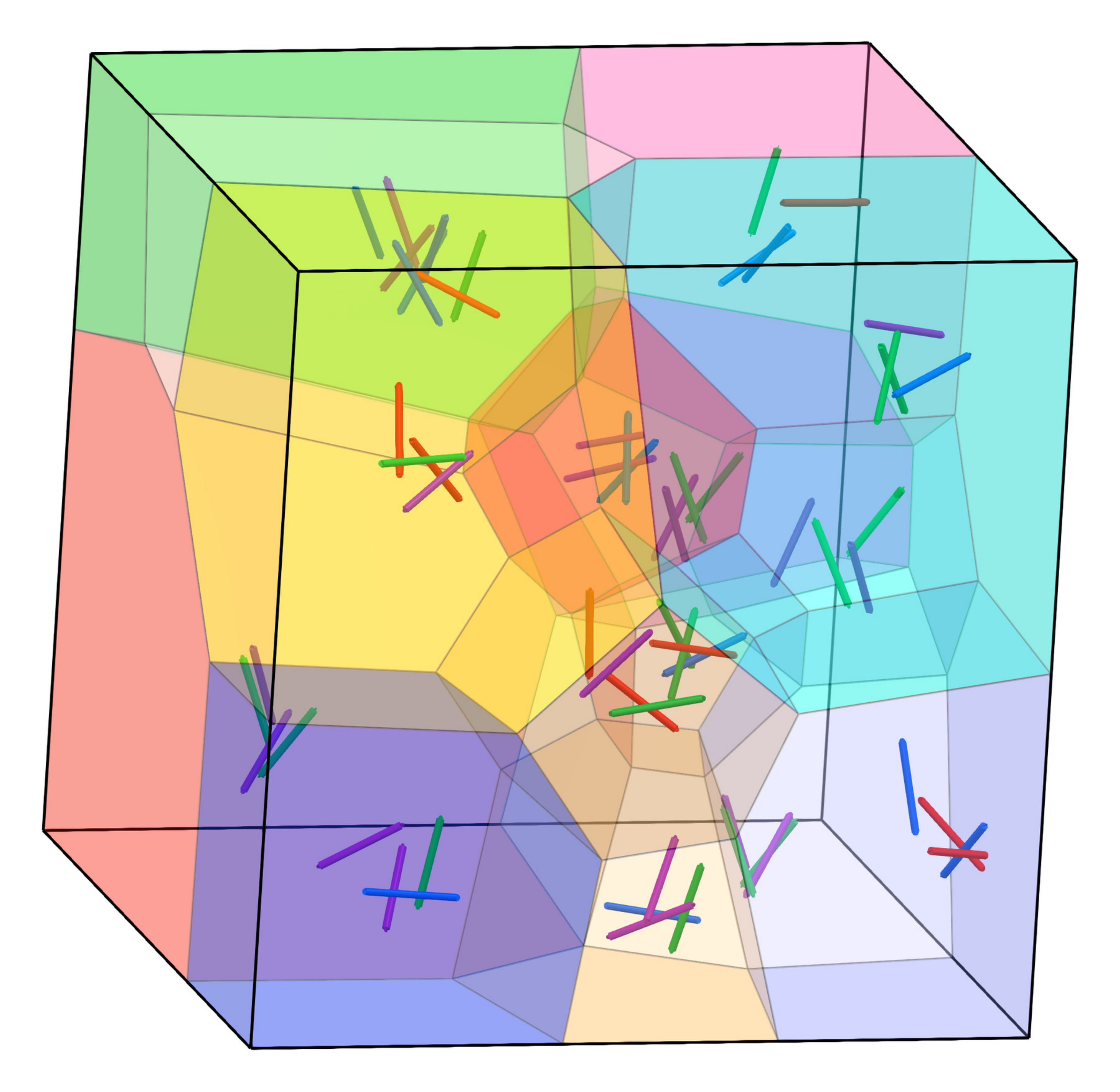}
        \caption{Initial source configuration.}
        \label{fig:ddd_step0}
    \end{subfigure}
    \hfill
    \begin{subfigure}[b]{0.48\textwidth}
        \centering
        \includegraphics[width=0.80\textwidth]{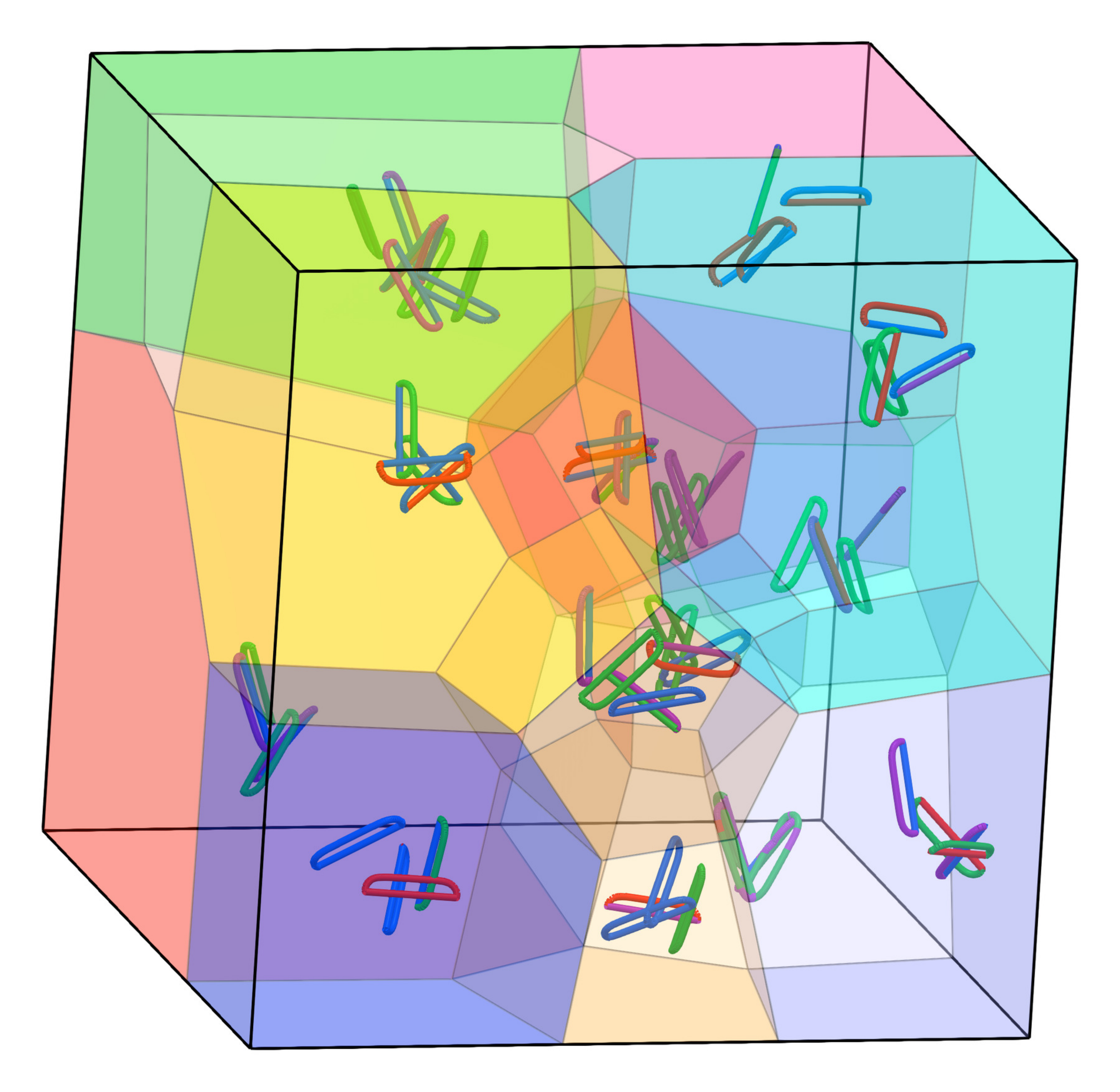}
        \caption{Early source bow-out.}
        \label{fig:ddd_step75}
    \end{subfigure}

    \vspace{0.25cm}

    \begin{subfigure}[b]{0.48\textwidth}
        \centering
        \includegraphics[width=0.80\textwidth]{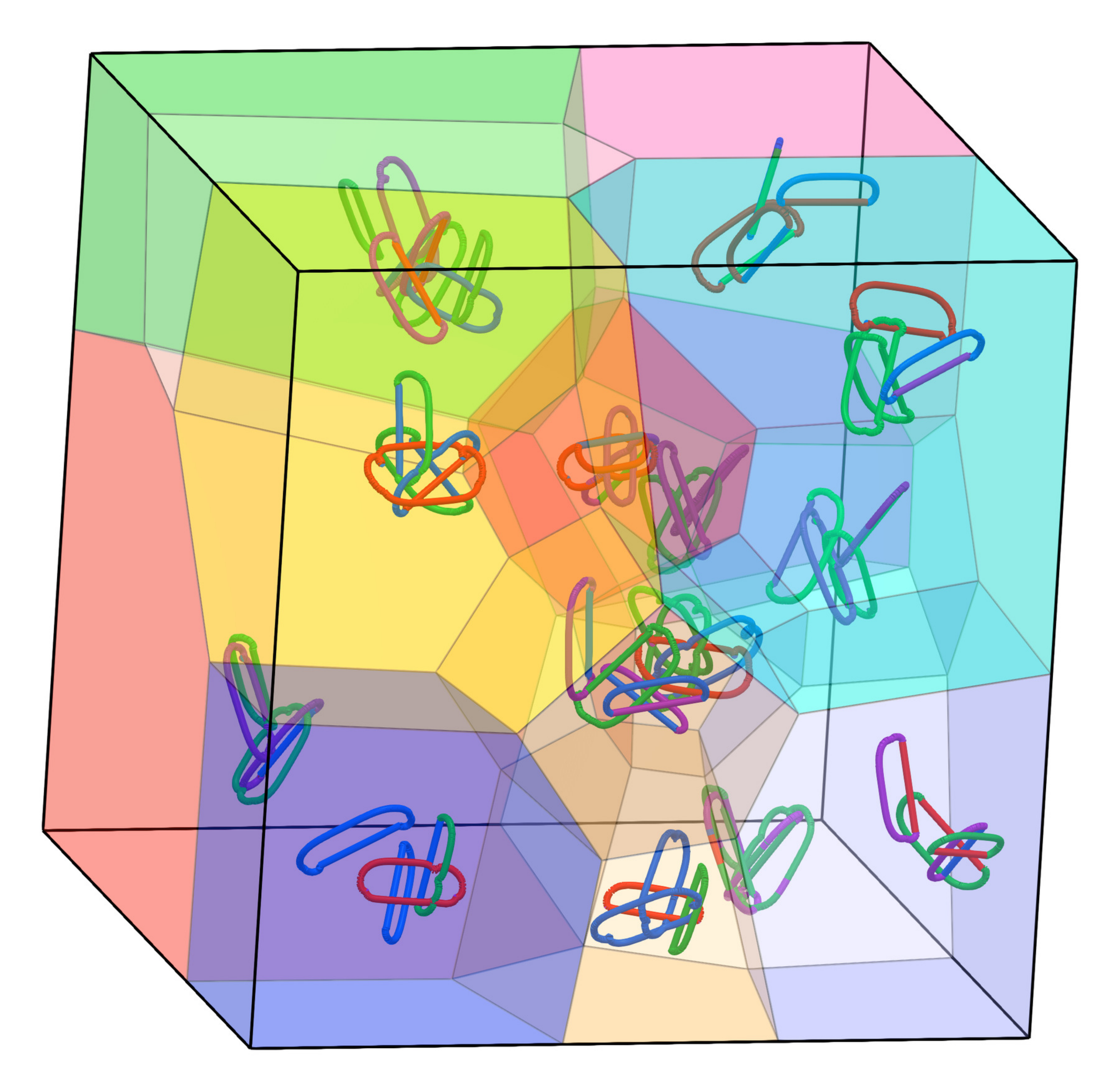}
        \caption{Loop expansion.}
        \label{fig:ddd_step150}
    \end{subfigure}
    \hfill
    \begin{subfigure}[b]{0.48\textwidth}
        \centering
        \includegraphics[width=0.80\textwidth]{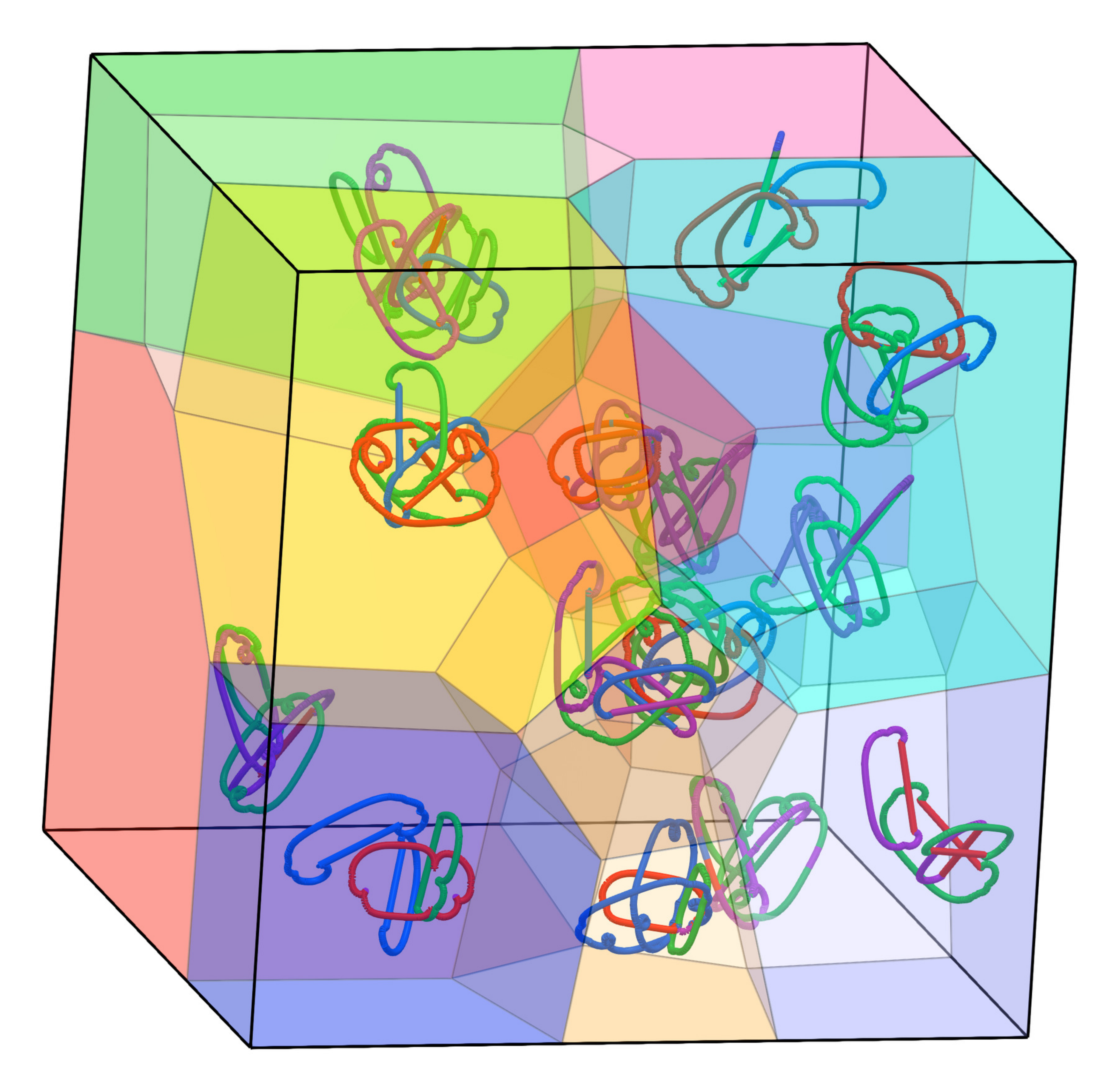}
        \caption{Developed dislocation network.}
        \label{fig:ddd_step225}
    \end{subfigure}

    \vspace{0.15cm}
    \includegraphics[width=0.92\textwidth]{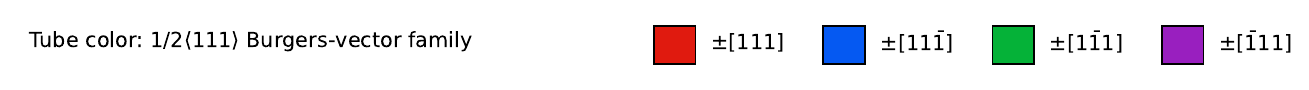}

    \caption{Evolution of the representative 15-grain DDD configuration. The prescribed five-node Frank--Read-type source configurations bow out and form extended loop-like structures under the applied stress. Tube colors denote \(\frac{1}{2}\langle111\rangle\) Burgers-vector families, and transparent colored regions identify grains. The visual sequence demonstrates source-like bow-out and heterogeneous network development but is not a continuation of a specific MD trajectory.}
    \label{fig:ddd_evolution}
\end{figure}

The initially compact source segments bow out under the applied stress and develop increasingly extended loop-like configurations (Fig.~\ref{fig:ddd_evolution}). The network evolution is heterogeneous among grains because source orientations and resolved driving forces vary throughout the aggregate. Some grains develop dense, strongly curved networks, whereas others remain comparatively sparse. The sequence demonstrates that an initialized \(\frac{1}{2}\langle111\rangle\) source population can evolve collectively in a crystallographically heterogeneous tungsten polycrystal.

For this representative calculation, a density proxy was evaluated from the segment count,
\begin{equation}
\rho_{\mathrm{seg}}
=
\frac{N_{\mathrm{seg}}\overline{L}_{\mathrm{seg}}}{V},
\qquad
\overline{L}_{\mathrm{seg}}=15b,
\label{eq:ddd_density_proxy}
\end{equation}
where \(N_{\mathrm{seg}}\) is the number of discretized segments, \(\overline{L}_{\mathrm{seg}}\) is the assumed mean segment length, and \(V\) is the DDD volume. Because dynamic remeshing also changes \(N_{\mathrm{seg}}\), \(\rho_{\mathrm{seg}}\) is used as a network-development proxy rather than an exact line-length density.

\begin{figure}[htbp]
    \centering
    \includegraphics[width=\linewidth]{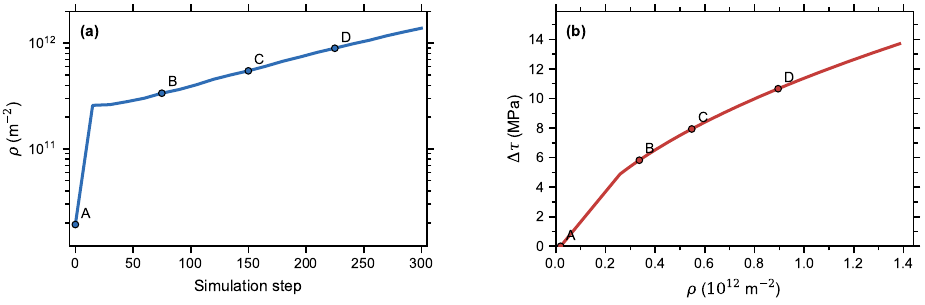}
    \caption{Quantitative descriptors of the representative DDD calculation. (a) Segment-count-based density proxy, \(\rho_{\mathrm{seg}}\), as a function of simulation step. States A--D correspond to Fig.~\ref{fig:ddd_evolution}. (b) Taylor-type resistance increment derived from \(\rho_{\mathrm{seg}}\). The resistance increment is a post-processed indicator and is not an independently calculated macroscopic flow stress.}
    \label{fig:ddd_density_hardening}
\end{figure}

The density proxy rises from approximately \(1.93\times10^{10}\) to \(1.38\times10^{12}~\mathrm{m}^{-2}\), corresponding to an increase by a factor of approximately 72. A Taylor-type resistance was then calculated from
\begin{equation}
\tau_{\mathrm{T}}
=
\alpha\mu b\sqrt{\rho_{\mathrm{seg}}},
\qquad
\Delta\tau_{\mathrm{T}}
=
\tau_{\mathrm{T}}-\tau_{\mathrm{T},0},
\label{eq:ddd_taylor_indicator}
\end{equation}
using \(\alpha=0.3\), \(\mu=160.6~\mathrm{GPa}\), and \(b=2.741\times10^{-10}~\mathrm{m}\). The resulting monotonic increase in \(\Delta\tau_{\mathrm{T}}\) follows directly from the assumed \(\sqrt{\rho_{\mathrm{seg}}}\) relation. Fig.~\ref{fig:ddd_density_hardening} therefore quantifies the growth of the network proxy and its associated Taylor-type resistance; it does not constitute an independently predicted stress--strain hardening curve.

The DDD calculation addresses a different question from the MD simulations. MD resolves crack-tip structural rearrangement, dislocation nucleation, and mechanical instability over nanometer and sub-nanosecond scales. The representative DDD calculation begins from prescribed five-node Frank--Read-type configurations in a crack-free micrometer-scale polycrystal and illustrates how this initialized population can bow out and develop into a heterogeneous network. It therefore supplies mesoscale context for collective dislocation evolution, but it is not a sequential continuation of the MD trajectories. The heterogeneous network in Fig.~\ref{fig:ddd_evolution} is consistent with the general expectation that crystallographic orientation and source geometry produce grain-dependent activity under a nominally uniform load \cite{bulatov2006,arsenlis2007,weygand2002}. DDD has previously been used to analyze the competition between dislocation activity and crack-tip loading in tungsten \cite{tarleton2009ddd}; however, the present DDD geometry contains no crack and should not be interpreted as a crack-shielding calculation. Likewise, it contains no Re chemistry and cannot validate the atomistic Re mechanism.

The quantitative DDD measures require an additional qualification. The reported \(\rho_{\mathrm{seg}}\) is based on segment count and an assumed mean segment length, so part of its evolution can be influenced by dynamic remeshing. The Taylor-type resistance is then algebraically derived from this proxy. In addition, the low-angle term in the heuristic boundary-load correction is inactive in the reported trajectory because no segment meets its proximity criterion at the recorded states. The robust DDD evidence is therefore the visual bow-out of the prescribed configurations and development of a heterogeneous network; the density and resistance curves are compact descriptors of that evolution rather than calibrated predictions of physical dislocation density or macroscopic hardening. Direct scale bridging would require actual line-length tracking together with atomistically informed mobility, nucleation, solute, obstacle, and interface parameters.

\section{Conclusion}
\label{sec:conclusion}
The atomistic calculations establish that heating the edge-cracked pure-W specimen primarily produces thermal weakening under the present high-rate loading protocol. The instability stress, effective initial stiffness, and volumetric work density to instability decrease overall with temperature, whereas the instability strain remains scattered and is often lower rather than systematically delayed. The fitted work-loss midpoint, \(T_{\mathrm{MD}}^{*}\approx870~\mathrm{K}\), has a broad 255-K width and should therefore be used only as an internal marker separating regimes of the simulated response. It is not a direct estimate of a macroscopic or experimental DBTT.

The Re effect is expressed more consistently through deformation accumulated up to the operational instability than through peak strength. W--10Re reaches the global processed-stress maximum within \(\epsilon\leq0.05\) at a larger strain than W at 15 of 16 temperatures and accumulates more pre-instability work density at 13 of 16 temperatures, while the peak-stress difference changes sign. At \(\epsilon=0.045\) and 1000~K, the retained \(\frac{1}{2}\langle111\rangle\) edge fraction increases from 0.38 in W to 0.67 in W--10Re. The common-window near-tip response is temperature dependent: the alloy has lower peak line length at 700, 900, and 1000~K, whereas both its peak and time-averaged values are higher at 1200~K. These linked observations are consistent with Re altering defect-mediated crack-tip accommodation, but retained line character alone does not establish a mobility mechanism.

The separate DDD calculation illustrates source bow-out and heterogeneous dislocation-network development in crack-free polycrystalline W. Because the MD results are based on one geometry and one trajectory per condition, the conclusions remain comparative and specific to the present simulation protocol. Future work should include replicate simulations, direct measurements of crack growth and blunting, and quantitative coupling between the MD and DDD models. Within these limits, delayed instability and edge-enriched retained plasticity, rather than uniform strengthening, are the principal signatures of the representative W--Re response.

\section*{Data and Code Availability}

The processed data and analysis scripts supporting the findings of this study
are available from the corresponding author upon reasonable request.

\section*{Acknowledgements}
This work was supported by the Department of Mechanical and Electrical Engineering at Merrimack College. The authors acknowledge the use of computational resources at the Massachusetts Green High Performance Computing Center (MGHPCC). This research also benefited from high performance computing allocations provided by the National Science Foundation through ACCESS (awards MAT250103). Additional computational resources were supported by Argonne National Laboratory under the Director's Discretionary allocation for the project \textit{GNNMD}. Further support was provided through a National Science Foundation MRI Award to Wilkes University (Award No.\ 1920129), which contributed essential computational infrastructure for this study.

\bibliographystyle{unsrt}
\bibliography{references}

\end{document}